\documentclass[aps,prd,preprint,superscriptaddress,preprintnumbers,nofootinbib,tightenlines,floatfix]{revtex4-1}

\newcommand{\usepreprint}{1}

\usepackage{graphicx} 
\usepackage{dcolumn}  
\usepackage{bm}       
\usepackage{amsmath}       
\usepackage{ifthen}
\usepackage{rotating}

\newcommand{\etal}{{\it et al.}}

\newcommand{\rar}{\rightarrow}

\newcommand{\pip}{\pi^+}
\newcommand{\pim}{\pi^-}
\newcommand{\pio}{\pi^0}

\newcommand{\eorep}{\eta^{(}{'}^{)}}

\newcommand{\chin}{\chi_{c1}}

\newcommand{\psichin}{\psi(2S)\rar\gamma\chin}
\newcommand{\chine}{\chin\rar\eta\pip\pim}
\newcommand{\chinep}{\chin\rar\eta'\pip\pim}

\newcommand{\spp}{S_{\pi\pi}}
\newcommand{\sppz}{S^{0}_{\pi\pi}}
\newcommand{\sppo}{S^{1}_{\pi\pi}}
\newcommand{\skk}{S_{KK}}
\newcommand{\aze}{a_{0}(980)}

\newcommand{\atw}{a_{2}(1320)}
\newcommand{\ftw}{f_{2}(1270)}
\newcommand{\ffo}{f_{4}(2050)}

\newcommand{\pione}{\pi_{1}(1600)}

\newcommand{\mevcc}{\mathrm{MeV}/c^2}
\newcommand{\mevc}{\mathrm{MeV}/c}
\newcommand{\mev}{\mathrm{MeV}}
\newcommand{\gevcc}{\mathrm{GeV}/c^2}
\newcommand{\gevc}{\mathrm{GeV}/c}
\newcommand{\gev}{\mathrm{GeV}}

\newcommand{\cm}{\mathrm{cm}}

\newcommand{\etaprime}{\eta^{\prime}}

\newcommand{\decaya}{\chi_{c1}\to\eta\pip\pim}
\newcommand{\decayb}{\chi_{c1}\to\etaprime\pip\pim}
\newcommand{\fulldecaya}{\psi(2S)\to\gamma\chi_{c1};~\decaya}
\newcommand{\fulldecayb}{\psi(2S)\to\gamma\chi_{c1};~\decayb}

\newcommand{\chidof}{\chi^2/\mathrm{d.o.f.}}

\begin{document}

\preprint{CLNS 11/2080}  
\preprint{CLEO 11-06}    

\ifthenelse{\equal{\usepreprint}{1}}{
\title{\boldmath Amplitude analyses of the decays \\ $\chi_{c1}\rightarrow\eta\pi^+\pi^-$ and $\chi_{c1}\rightarrow\eta^{\prime}\pi^+\pi^-$}
}{
\title{\boldmath Amplitude analyses of the decays $\chi_{c1}\rightarrow\eta\pi^+\pi^-$ and $\chi_{c1}\rightarrow\eta^{\prime}\pi^+\pi^-$}
}

\author{G.~S.~Adams}
\author{J.~Napolitano}
\affiliation{Rensselaer Polytechnic Institute, Troy, New York 12180, USA}
\author{K.~M.~Ecklund}
\affiliation{Rice University, Houston, Texas 77005, USA}
\author{J.~Insler}
\author{H.~Muramatsu}
\author{C.~S.~Park}
\author{L.~J.~Pearson}
\author{E.~H.~Thorndike}
\affiliation{University of Rochester, Rochester, New York 14627, USA}
\author{S.~Ricciardi}
\affiliation{STFC Rutherford Appleton Laboratory, Chilton, Didcot, Oxfordshire, OX11 0QX, United Kingdom}
\author{C.~Thomas}
\affiliation{STFC Rutherford Appleton Laboratory, Chilton, Didcot, Oxfordshire, OX11 0QX, United Kingdom}
\affiliation{University of Oxford, Oxford OX1 3RH, United Kingdom}
\author{M.~Artuso}
\author{S.~Blusk}
\author{R.~Mountain}
\author{T.~Skwarnicki}
\author{S.~Stone}
\author{L.~M.~Zhang}
\affiliation{Syracuse University, Syracuse, New York 13244, USA}
\author{G.~Bonvicini}
\author{D.~Cinabro}
\author{A.~Lincoln}
\author{M.~J.~Smith}
\author{P.~Zhou}
\author{J.~Zhu}
\affiliation{Wayne State University, Detroit, Michigan 48202, USA}
\author{P.~Naik}
\author{J.~Rademacker}
\affiliation{University of Bristol, Bristol BS8 1TL, United Kingdom}
\author{D.~M.~Asner}
\altaffiliation[Present address: ]{Pacific Northwest National Laboratory, Richland, WA 99352}
\author{K.~W.~Edwards}
\author{K.~Randrianarivony}
\author{G.~Tatishvili}
\altaffiliation[Present address: ]{Pacific Northwest National Laboratory, Richland, WA 99352}
\affiliation{Carleton University, Ottawa, Ontario, Canada K1S 5B6}
\author{R.~A.~Briere}
\author{H.~Vogel}
\affiliation{Carnegie Mellon University, Pittsburgh, Pennsylvania 15213, USA}
\author{P.~U.~E.~Onyisi}
\author{J.~L.~Rosner}
\affiliation{University of Chicago, Chicago, Illinois 60637, USA}
\author{J.~P.~Alexander}
\author{D.~G.~Cassel}
\author{S.~Das}
\author{R.~Ehrlich}
\author{L.~Gibbons}
\author{S.~W.~Gray}
\author{D.~L.~Hartill}
\author{B.~K.~Heltsley}
\author{D.~L.~Kreinick}
\author{V.~E.~Kuznetsov}
\author{J.~R.~Patterson}
\author{D.~Peterson}
\author{D.~Riley}
\author{A.~Ryd}
\author{A.~J.~Sadoff}
\author{X.~Shi}
\author{W.~M.~Sun}
\affiliation{Cornell University, Ithaca, New York 14853, USA}
\author{J.~Yelton}
\affiliation{University of Florida, Gainesville, Florida 32611, USA}
\author{P.~Rubin}
\affiliation{George Mason University, Fairfax, Virginia 22030, USA}
\author{N.~Lowrey}
\author{S.~Mehrabyan}
\author{M.~Selen}
\author{J.~Wiss}
\affiliation{University of Illinois, Urbana-Champaign, Illinois 61801, USA}
\author{J.~Libby}
\affiliation{Indian Institute of Technology Madras, Chennai, Tamil Nadu 600036, India}
\author{M.~Kornicer}
\author{R.~E.~Mitchell}
\author{M.~R.~Shepherd}
\author{A.~Szczepaniak}
\affiliation{Indiana University, Bloomington, Indiana 47405, USA }
\author{D.~Besson}
\affiliation{University of Kansas, Lawrence, Kansas 66045, USA}
\author{T.~K.~Pedlar}
\affiliation{Luther College, Decorah, Iowa 52101, USA}
\author{D.~Cronin-Hennessy}
\author{J.~Hietala}
\affiliation{University of Minnesota, Minneapolis, Minnesota 55455, USA}
\author{S.~Dobbs}
\author{Z.~Metreveli}
\author{K.~K.~Seth}
\author{A.~Tomaradze}
\author{T.~Xiao}
\affiliation{Northwestern University, Evanston, Illinois 60208, USA}
\author{L.~Martin}
\author{A.~Powell}
\author{G.~Wilkinson}
\affiliation{University of Oxford, Oxford OX1 3RH, United Kingdom}
\author{J.~Y.~Ge}
\author{D.~H.~Miller}
\author{I.~P.~J.~Shipsey}
\author{B.~Xin}
\affiliation{Purdue University, West Lafayette, Indiana 47907, USA}
\collaboration{CLEO Collaboration}
\noaffiliation


\date{September 25, 2011}

\begin{abstract} 
Using a data sample of $2.59\times10^7$~$\psi(2S)$ decays obtained with the CLEO-c detector, we perform amplitude analyses of the complementary decay chains $\fulldecaya$ and $\fulldecayb$.  We find evidence for an exotic $P$-wave $\eta^\prime\pi$ amplitude, which, if interpreted as a resonance, would have parameters consistent with the $\pi_1(1600)$ state reported in other production mechanisms.  We also make the first observation of the decay $a_0(980)\to\etaprime\pi$ and measure the ratio of branching fractions ${\cal B}(a_0(980)\to\etaprime\pi)/{\cal B}(a_0(980)\to\eta\pi) = 0.064\pm0.014\pm 0.014$.  The $\pi\pi$ spectrum produced with a recoiling $\eta$ is compared to that with $\etaprime$ recoil. 
\end{abstract}


\maketitle


\section{\label{intro}Introduction}

Hadronic charmonium decays, in which charm and anti-charm quarks annihilate into gluons, provide an excellent opportunity to study light mesons.  The combination of well-defined initial states and the availability of a wide variety of final states allows for the strategic selection of reactions to isolate and study different light meson systems. The decays $\decaya$ and $\decayb$, in particular, have two interesting characteristics.  

First, since the quark and SU(3) flavor of the $\eta$ and $\etaprime$ are relatively well-known, one could, in principle, model these decays using what is known about the OZI rule and SU(3) flavor symmetry and learn about the $\pi\pi$ isoscalar states recoiling against them.  Such a technique has been proposed for other $\chi_{cJ}$ decay channels~\cite{OZI}.  

Second, the decays $\decaya$ and $\decayb$ provide an opportunity to search for exotic $J^{PC}=1^{-+}$ states in the $\eta\pi$ and $\etaprime\pi$ systems.  In fact, the only two-body $S$-wave decays of the $\chi_{c1}$ available in these channels necessarily have the $\eta\pi$ or $\etaprime\pi$ system in a configuration with $J^{PC}=1^{-+}$.  Two exotic candidates, the $\pi_1(1400)$ and the $\pi_1(1600)$, have been reported in other production mechanisms to have decays to $\eta\pi$ and $\etaprime\pi$, respectively.  The $\pi_1(1400)$ has been reported primarily decaying to $\eta\pi$~\cite{GAMS88,KEK93,E852epp,CBAR98}, while the $\pi_1(1600)$ has been reported to decay to $\etaprime\pi$~\cite{E85201,VES93}, $b_1\pi$~\cite{baker,lu}, $f_1\pi$~\cite{kuhn}, and $\rho\pi$~\cite{alekseev}.

We present amplitude analyses of the processes $\fulldecaya$ and $\fulldecayb$ in which we study the various $\eta\pi$ and $\pi\pi$ resonances produced in the decays of the $\chi_{c1}$.  In Section~\ref{DataSelect}, we describe the data-selection process that results in a sample of $\chi_{c1}$ decays with estimated backgrounds below $5\%$.  In Section~\ref{AmpAna}, we describe our construction of amplitudes using the helicity formalism.  Here we assume that the $\decaya$ and $\decayb$ decays proceed through a sequence of two-body decays, where the intermediate states have well-defined quantum numbers.  We pay special attention to the treatment of the $\pi\pi$ $S$-wave, which utilizes independent experimental data on $\pi\pi$ $S$-wave scattering~\cite{LEON9499}.  We also describe our fitting procedure, based on the unbinned extended maximum-likelihood method.

The highlights of our analysis, detailed in Section~\ref{Results}, are
\begin{itemize}
 \item evidence for a $P$-wave $\eta^\prime\pi$ amplitude, which, when parametrized by an exotic $J^{PC} = 1^{-+}$ resonance, has properties consistent with those of the $\pi_1(1600)$ reported in other production mechanisms;
 \item the first direct observation of the decay $a_0(980)\to\etaprime\pi$, a measurement of the ratio of branching fractions ${\cal B}(a_0(980)\to\etaprime\pi)/{\cal B}(a_0(980)\to\eta\pi)$, and a characterization of the $a_0(980)$ lineshape; and
 \item the observation of qualitative differences in the $\pi\pi$ system when it is produced against the $\eta$ or $\etaprime$. 
 \end{itemize}
Finally, in Section~\ref{Systematics}, we evaluate and discuss systematic errors.

\section{\label{DataSelect}Data selection}

We select candidate events of the form $\fulldecaya$ and $\fulldecayb$ using $25.9\times10^6$ $\psi(2S)$ decays collected by the CLEO-c detector
at the Cornell Electron Storage Ring. We reconstruct the $\eta$ ($\etaprime$) in three (six) different decay topologies comprising $94.6\pm0.7\%$ ($83.8\pm1.8\%$) of its total decay width (Table~\ref{tab:fstates}).  We then select the $\chi_{c1}$ using the energy of the photon from $\psi(2S)\to\gamma\chi_{c1}$.

Final state photons and charged pions are measured by the CLEO-c detector~\cite{Kubota:1992ww}, which covers a solid angle of $93\%$. 
The detector has a 1~Tesla superconducting magnet enclosing 
two drift chambers and a ring imaging Cherenkov (RICH) system 
for tracking charged particles and particle identification.
Enclosed inside the solenoid are also a barrel and two endcap 
CsI-crystal calorimeters. 
The energy resolution for $100~\mev$ $(1~\gev)$ photons is $5.0\%$ $(2.2\%)$,
while the momentum resolution for charged tracks in the drift chambers is $0.6\%$ at $1~\gevc$. 

Charged tracks are required to have momentum $p>18.4~\mevc$
and originate within a cylindrical volume, with $20~\cm$ length and $2~\cm$ radius, centered around the interaction point.
The $\pi^{\pm}$ candidates are then required to have ionization losses ($dE/dx$) within $3\sigma$ of those expected for charged pions. Photons reconstructed inside the calorimeters, with polar angles 
$|\cos\theta| < 0.79$ and $0.85 < |\cos\theta| < 0.93$,  
are required to have energy $E>20~\mev$ and separation from charged tracks.
Two-photon four-momenta are kinematically constrained  to select the $\pio\to\gamma\gamma$ and
$\eta\to\gamma\gamma$ candidates, 
with a requirement that the respective invariant mass is within 3$\sigma$ of the $\pio$ or $\eta$ rest mass. Finally, the total four-momentum of all of the final state particles of a given topology is kinematically constrained to the initial $\psi(2S)$ four-momentum and the $\chi^2$ of the resulting fit is required to satisfy $\chidof<5$.  If multiple combinations of tracks and showers within an event pass all of these selection requirements (which occurs for $1.6\%$ of all selected events), only the combination with the best $\chidof$ is accepted. 

\begin{table}
\caption{\label{tab:fstates}
The decay modes of the $\eta$ and $\etaprime$ that are used to reconstruct $\fulldecaya$ and $\fulldecayb$, the branching fractions $\mathcal{B}$ of each decay mode~\cite{PDG10}, and the final state topology reconstructed with the detector.
}
\medskip
\begin{center}
\begin{tabular}{lrc}
\hline\hline
 $\eta^{(\prime)}$ Decay Mode  & ~~$\mathcal{B}$ [$\%$]~~ & ~~Final State~~\\
\hline 
  $\eta\to\gamma\gamma$ & 
39.3$\pm$0.2 &  
$3\gamma~\!1(\pip\pim)$ \\ 
  $\eta\to\pip\pim\pio$ & 
22.7$\pm$0.3 &  
$3\gamma~\!2(\pip\pim)$ \\
  $\eta\to\pio\pio\pio$ & 
32.6$\pm$0.2 &  
$7\gamma~\!1(\pip\pim)$ \\
\hline 
  $\etaprime\to\pip\pim\eta;~\eta\to\gamma\gamma$ & 
17.1$\pm$0.3 & 
$3\gamma~\!2(\pip\pim)$ \\ 
  $\etaprime\to\pip\pim\eta;~\eta\to\pip\pim\pio$ & 
9.9$\pm$0.2 & 
$3\gamma~\!3(\pip\pim)$ \\ 
  $\etaprime\to\pip\pim\eta;~\eta\to\pio\pio\pio$ & 
14.1$\pm$0.2 & 
$7\gamma~\!2(\pip\pim)$ \\  
  $\etaprime\to\gamma\pip\pim$                    & 
29.3$\pm$0.6 & 
$2\gamma~\!2(\pip\pim)$ \\ 
  $\etaprime\to\pio\pio\eta;~\eta\to\gamma\gamma$ &  
8.5$\pm$0.3 & 
$7\gamma~\!1(\pip\pim)$ \\ 
  $\etaprime\to\pio\pio\eta;~\eta\to\pip\pim\pio$ &  
4.9$\pm$0.2 & 
$7\gamma~\!2(\pip\pim)$ \\ 
\hline \hline
\end{tabular}
\end{center}
\end{table}

\begin{figure}
\ifthenelse{\equal{\usepreprint}{1}}{
\includegraphics[width=0.40\linewidth]{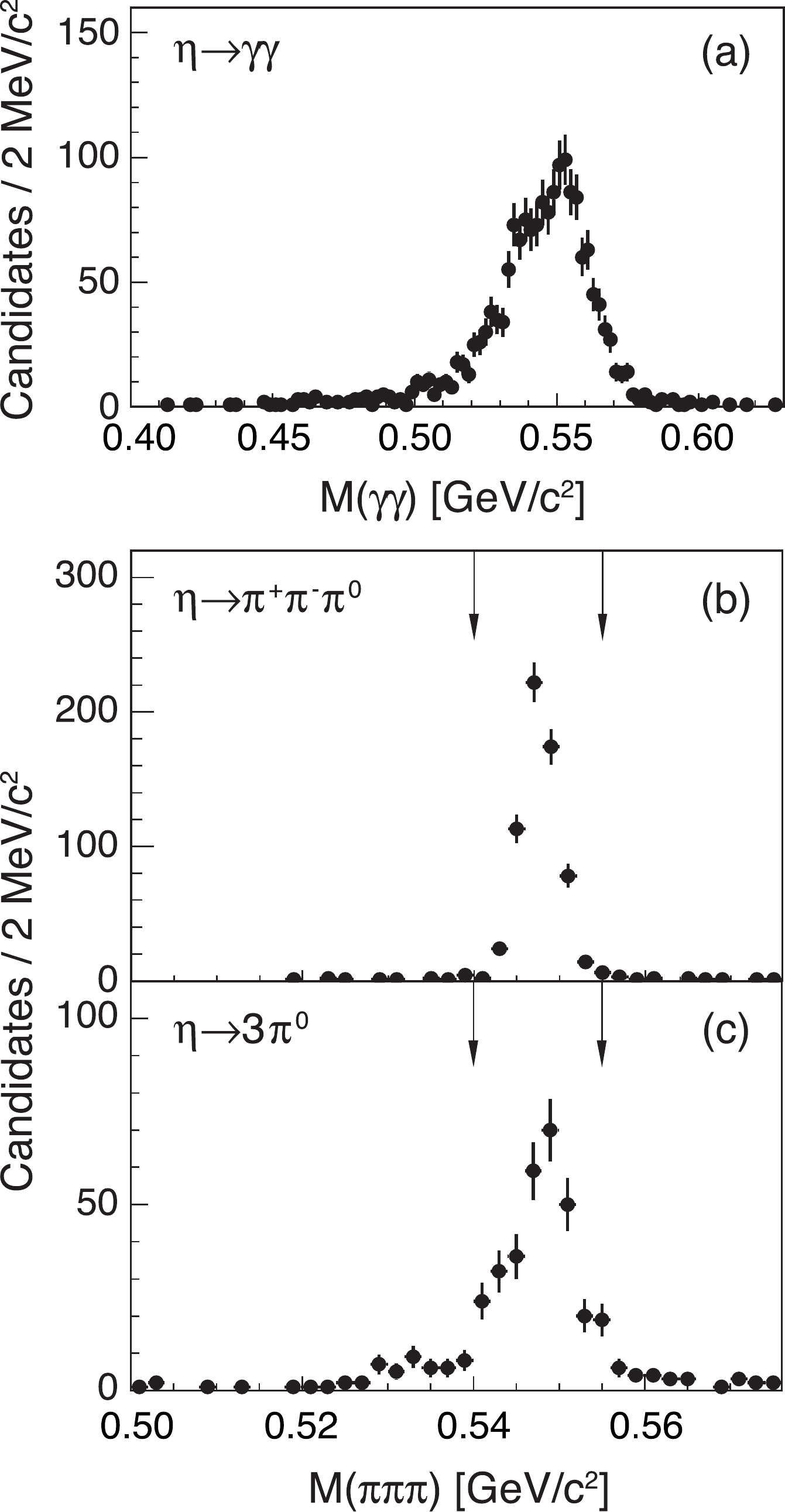}
}{
\includegraphics[width=0.75\linewidth]{4250911-008}
}
\caption{The invariant mass of $\eta$ candidates after selecting a $\chi_{c1}$ candidate and applying background suppression criteria.  Candidates in (a) are selected by requiring individual photon pairs be no more than 3$\sigma$ from the nominal $\eta$ mass; the arrows in (b) and (c) indicate the region used to select the $\eta$ candidates.
\label{fig:mEta}
}
\end{figure}

\begin{figure}
\ifthenelse{\equal{\usepreprint}{1}}{
\includegraphics[width=0.40\linewidth]{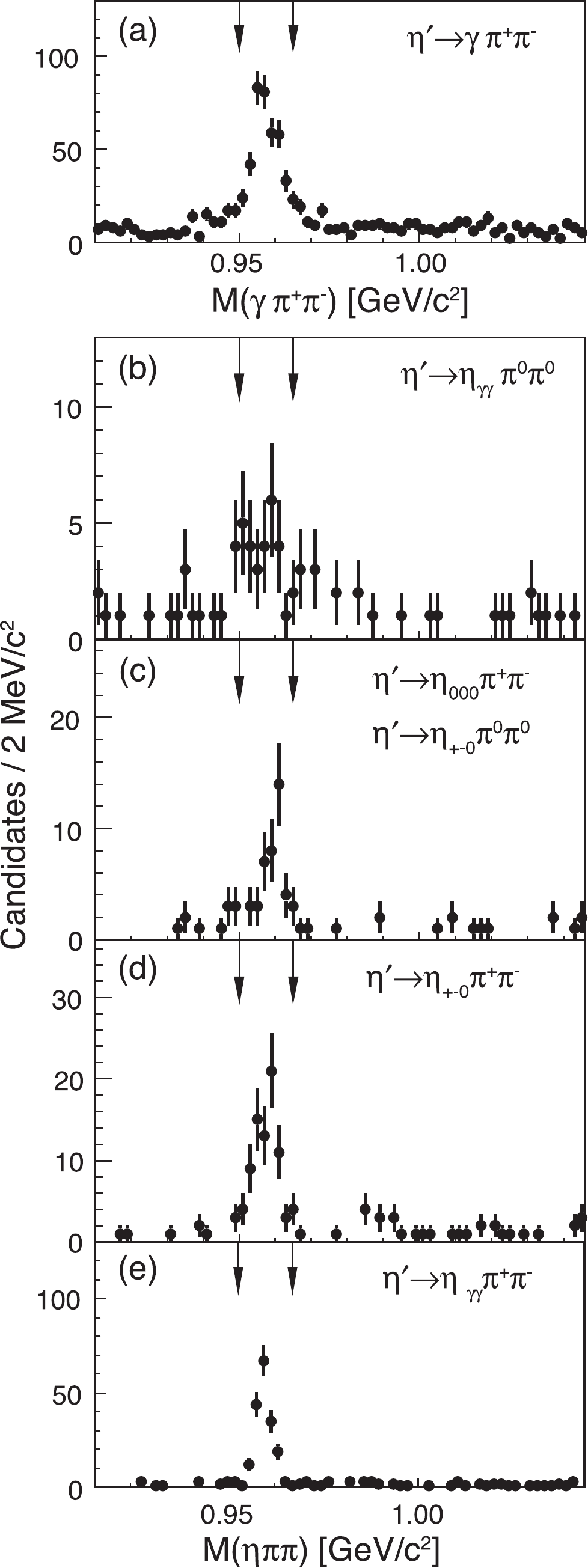}
}{
\includegraphics[width=0.75\linewidth]{4250911-011}
}
\caption{The invariant mass of $\etaprime$ candidates after selecting a $\chi_{c1}$ candidate and applying background suppression criteria. The arrows indicate the region used to select the $\eta^\prime$ candidates.
\label{fig:mEtap}
}
\end{figure}

To select the decays $\eta\to\pip\pim\pio$ and $\eta\to\pio\pio\pio$, the invariant mass of the three pions must be between $540$ and $555~\mevcc$ (see Fig.~\ref{fig:mEta}).  Similarly, to select the $\etaprime$ in its various decay modes the invariant mass of its decay products must fall between $950$ and $965~\mevcc$ (see Fig.~\ref{fig:mEtap}).  If multiple combinations of particles within an event can be used to form an $\eta$ (occurring in $0.9\%$ of selected events), one of these combinations is chosen randomly.  If there are multiple $\etaprime$ candidates (occurring in $1.4\%$ of selected events), the event is discarded.  Since the decays $\etaprime\to\pip\pim\eta;~\eta\to\pio\pio\pio$ and $\etaprime\to\pio\pio\eta;~\eta\to\pip\pim\pio$ share the same final state topology, no requirement is made on the internal $\eta$ mass.  

Specific backgrounds are further suppressed based on studies using a Monte Carlo (MC) sample of inclusive $\psi(2S)$ decays.  For the $\fulldecaya$ decay chain, the dominant backgrounds are due to $\psi(2S)\to\eta J/\psi$ and $\psi(2S)\to\gamma\chi_{c1};~\chi_{c1}\to\gamma J/\psi$, where $J/\psi\to\mu^+\mu^-$.  The first of these is suppressed by requiring the mass recoiling against the $\eta$ be separated from the $J/\psi$ mass by at least $20~\mevcc$.  The second is only a background for the $\eta\to\gamma\gamma$ mode and is similarly suppressed using the masses recoiling against the $\gamma\gamma$ combinations, which are required to be more than $35~\mevcc$ from the $J/\psi$ mass.

The largest backgrounds in the $\fulldecayb$ decay chain occur in the $\etaprime\to\gamma\pip\pim$ mode.  We suppress $J/\psi$ backgrounds, as above, by requiring the masses recoiling against the $\gamma\gamma$ and $\pip\pim$ systems to be more than $20~\mevcc$ away from the $J/\psi$ mass.  In addition, we treat $\pip\pim$ combinations as $\mu^+\mu^-$ and require their invariant masses be more than $15~\mevcc$ away from the $J/\psi$ mass.  There is also a substantial background from $\psi(2S)\to\pio2(\pip\pim)$, which we reduce by requiring that no two showers in an event be consistent with the $\pio$ mass within $3\sigma$.  We also enhance the signal to background for the $\etaprime\to\gamma\pip\pim$ mode by making a loose requirement that the $\pip\pim$ invariant mass be between $335$ and $895~\mevcc$, which is motivated by the apparent $\rho$ dominance in the $\pip\pim$ system. 

One additional background for the $\fulldecayb$ decay chain with $\etaprime\to\gamma\pip\pim$ is from $\psi(2S)\to\gamma\chi_{c0};~\chi_{c0}\to2(\pip\pim)$ where the radiated photon converts to an $e^+e^-$ pair outside the tracking region.  This is suppressed by requiring that the total energy of the two resulting showers is not consistent with the energy of the photon from $\psi(2S)\to\gamma\chi_{c0}$, {\it i.e.,} not between $225$ and  $295~\mev$, and that the cosine of the angle between the two showers is less than 0.97.  

\begin{figure*}
\includegraphics[width=0.7\linewidth]{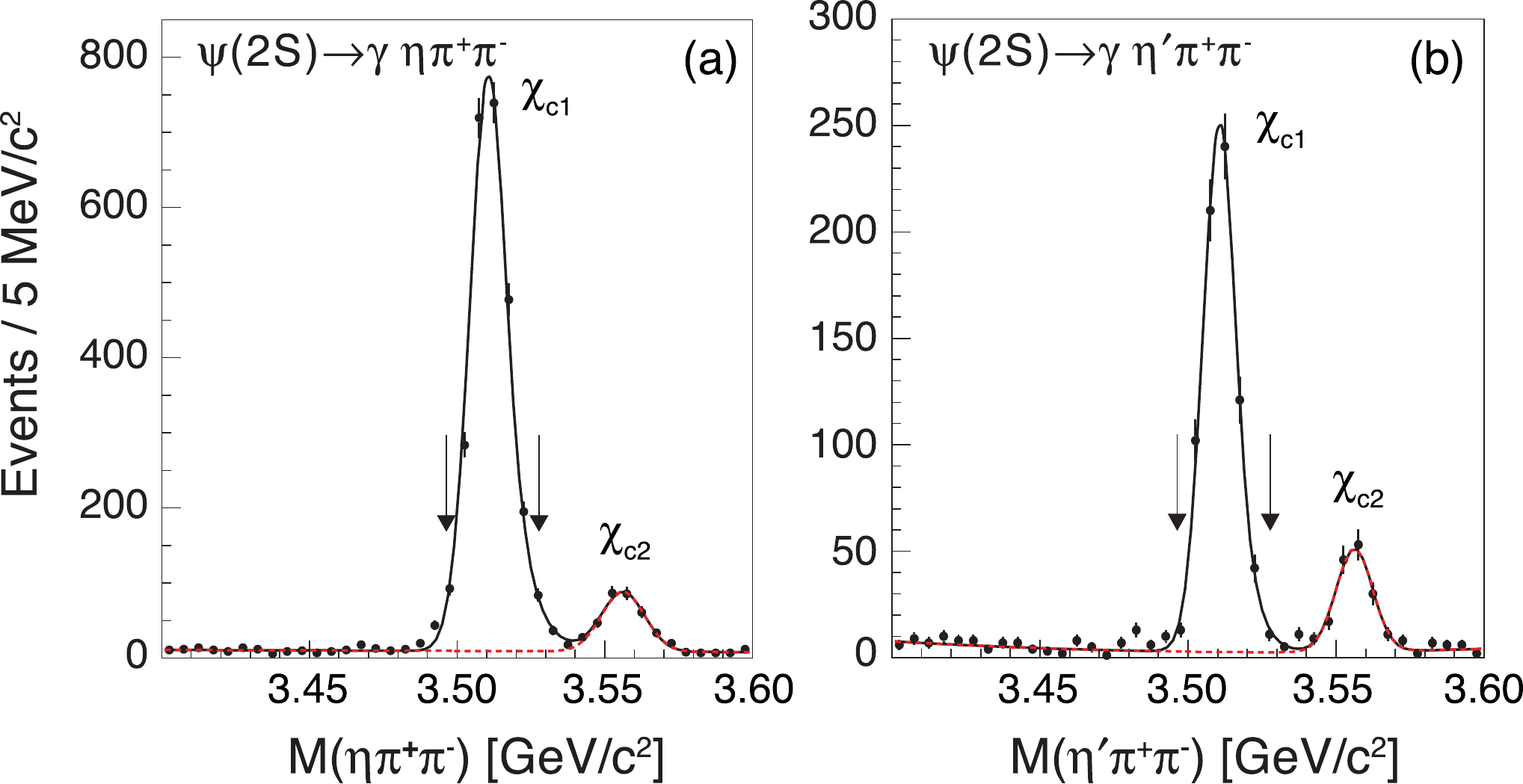}
\caption{The invariant mass distributions of the (a) $\eta\pip\pim$ and (b) $\etaprime\pip\pim$ 
candidates from selected $\psi(2S)\to\gamma\eta\pip\pim$ and $\psi(2S)\to\gamma\etaprime\pip\pim$ decays, respectively, after all background suppression criteria have been applied. The solid arrows indicate the regions used to
select the $\chi_{c1}$ signals.
\label{fig:mChic1}
}
\end{figure*}

Figure~\ref{fig:mChic1} shows the invariant mass distributions
of (a)~the $\eta\pip\pim$ and (b)~the $\etaprime\pip\pim$ candidates after combining all of the decay modes of the $\eta$ and $\etaprime$.
We select the $\chi_{c1}$ by requiring that the energy of the photon radiated from the $\psi(2S)$ be between $155$ and $185~\mev$ (indicated by the arrows in Fig.~\ref{fig:mChic1}).  Our final data samples consist of 2498 and 698 events in the $\fulldecaya$ and $\fulldecayb$ decay chains, respectively.  The background is estimated by fitting the data in Fig.~\ref{fig:mChic1} using a reverse Crystal Ball shape~\cite{cball} to describe the signal.  The background and $\chi_{c2}$ peak are described by a second order polynomial and a double Gaussian, respectively.  Peaking backgrounds have been subtracted by fitting the $\chi_c$ candidate mass distribution in $\eta^{(\prime)}$ mass sidebands; such backgrounds are negligible in all cases except the $\eta'\to\gamma\pi^+\pi^-$ decay mode.  The estimated signal purity for the $\eta\pip\pim$ ($\etaprime\pip\pim$) decay channel is 97.5\% (94.6\%) with an uncertainty of 0.3\% (1.3\%).


\section{\label{AmpAna}Amplitude analysis}

We perform amplitude analyses to disentangle the substructure present in the $\decaya$ and $\decayb$ decays.  We assume that the three-hadron decays of the $\chin$ proceed through a sequence of two-body decays, where one participant is the ``isobar,'' a bound state of either $\eta^{(\prime)}\pi^\pm$ or $\pip\pim$ with total angular momentum $J$, and the other is a stable, non-interacting meson (the $\pi^{\mp}$ or $\eta^{(\prime)}$) produced with an orbital angular momentum $L$ with respect to the isobar.  All possible $\chi_{c1}$ decays through isobars with $J\leq4$ are listed in Table~\ref{tab:spins}.

\begin{table}
\caption{\label{tab:spins}
A list of $\chi_{c1}$ decay modes for all possible isobars with $J\leq4$.
}
\medskip
\begin{center}
\begin{tabular}{ccc}
\hline \hline
~~$\chi_{c1}$ Decay Mode~~ & $L$ & Isobar $J^{PC}$ \\
\hline
$a_0\pi;~a_0\to\eta^{(\prime)}\pi$       & $P$     & $0^{++}$ \\
$\pi_1\pi;~\pi_1\to\eta^{(\prime)}\pi$   & $S,D$   & $1^{-+}$ \\
$a_2\pi;~a_2\to\eta^{(\prime)}\pi$       & $P,F$   & $2^{++}$ \\
$a_4\pi;~a_4\to\eta^{(\prime)}\pi$       & $F,H$   & $4^{++}$ \\
$f_0\eta^{(\prime)};~f_0\to\pi\pi$       & $P$     & $0^{++}$ \\
$f_2\eta^{(\prime)};~f_2\to\pi\pi$       & $P,F$   & $2^{++}$ \\
$f_4\eta^{(\prime)};~f_4\to\pi\pi$       & $F,H$   & $4^{++}$ \\
\hline \hline
\end{tabular}
\end{center}
\end{table}

The general idea of an amplitude analysis is to fit the distribution of events observed with the detector to a coherent sum of physically-motivated amplitudes that describes the dynamics of the intermediate states.  We can define $I(\mathbf{x})$, the number of observed events per unit phase space, as
\begin{equation}
I(\mathbf{x}) = \sum_{M_\psi,\lambda_\gamma}\left|\sum_{\alpha}V_{M_\psi,\lambda_\gamma}^\alpha A_{M_\psi,\lambda_\gamma}^\alpha(\mathbf{x})\right|^2,
\label{eq:inten}
\end{equation}
where $\alpha$ indexes the $\chi_{c1}$ decay amplitudes and $M_\psi$ and $\lambda_\gamma$ index the polarization of the $\psi(2S)$ and the helicity of the photon, respectively.  We use $\mathbf{x}$ to denote a set of kinematic variables, {\it e.g.}, angles and invariant masses, that provide a complete description of the event.  The value of the decay amplitude at a location $\mathbf{x}$ in this multi-dimensional space is given by $A_{M_\psi,\lambda_\gamma}^\alpha(\mathbf{x})$.  The real fit parameters $V_{M_\psi,\lambda_\gamma}^\alpha$ determine the relative strengths of each $\chi_{c1}$ decay amplitude.

Section~\ref{sec:amplitudes} discusses the construction of the decay amplitudes used in the fit.  Section~\ref{sec:fitting} discusses the application of the extended maximum likelihood technique to this analysis in order to determine the optimal values of $V_{M_\psi,\lambda_\gamma}^\alpha$ that describe the data.

\subsection{\label{sec:amplitudes}Amplitude construction}

\subsubsection{General amplitude structure}

The amplitude for a given $\chi_{c1}$ decay mode $\alpha$ depends on the set of observed final state event kinematics $\mathbf{x}$, the assumed polarization of the initial state $\psi(2S)$, denoted $M_\psi$, and the helicity of the final state photon $\lambda_\gamma$.  The general form is given by
\begin{widetext}
\begin{eqnarray}\label{eq:Amp}
 A^{\alpha}_{M_\psi,\lambda_\gamma}(\mathbf{x}) & = & \sum_{\lambda_{\chi}=\pm1,0} 
           C(M_\psi,\lambda_\gamma,\lambda_\chi) 
           \sum_{M^\prime_{\chi}=\pm1,0} D^{1*}_{M^\prime_{\chi},-\lambda_{\chi}}(\phi_\gamma,\theta_\gamma,0) \times
       \nonumber \\
       & ~ & \sum_{M^\prime_L, M^\prime_J }
       \langle 1 M^\prime_\chi| LM^\prime_L, JM^\prime_J\rangle Y^{M^\prime_L*}_L(\theta^\prime_{I},\phi^\prime_{I}) 
       Y^{M^\prime_J*}_J(\theta^\prime_h,\phi^\prime_h) p^Lq^J
       T_{\alpha}(s),
\end{eqnarray}
\end{widetext}
where summations in the second line are performed over all possible values $M^\prime_L$ and $M^\prime_J$, the projections of $L$ and $J$, respectively.  We briefly provide a term-by-term description of this expression.

The first factor in Eq.~(\ref{eq:Amp}), $C(M_\psi,\lambda_\gamma,\lambda_\chi)$, is used to transform the helicity
amplitude for the radiative decay into the multipole basis.  The $\psi(2S) \rar \gamma \chin$ radiative transition
is dominated by the electric dipole ($E1$) transition, while the magnetic quadrupole ($M2$) transition contributes $\approx3\%$~\cite{CLEO506} of the total rate.  In our analysis, we use the $E1$ amplitude to derive our results and check the sensitivity of
the results to the presence of a small $M2$ amplitude.  The $E1$ or $M2$ amplitude can be constructed with the following choice of $C$:
\begin{widetext}
\begin{eqnarray}\label{eq:alpha} 
  C(M_\psi,\lambda_\gamma,\lambda_\chi)  & = &  \sqrt{\frac{3}{8\pi}} 
                D^1_{M_\psi,\lambda_\gamma-\lambda_\chi}(\phi_\gamma,\theta_\gamma,0) \times \nonumber \\
      &&  \begin{cases}
   \left ( \delta_{\lambda_{\gamma},1} \delta_{\lambda_{\chi},1}
           - \delta_{\lambda_{\gamma},-1} \delta_{\lambda_{\chi},-1}
           +( \delta_{\lambda_{\gamma},1} - \delta_{\lambda_{\gamma},-1} ) \delta_{\lambda_{\chi},0}
   \right )   & \text{for $E1$, or}     \\
   \left ( \delta_{\lambda_{\gamma},1}\delta_{\lambda_{\chi},1}
           - \delta_{\lambda_{\gamma},-1} \delta_{\lambda_{\chi},-1}
	  -( \delta_{\lambda_{\gamma},1} - \delta_{\lambda_{\gamma},-1} ) \delta_{\lambda_{\chi},0}  
   \right )      & \text{for $M2$.}
        \end{cases} 
\end{eqnarray}
\end{widetext}

\begin{figure*}
\begin{center}
\includegraphics[width=0.7\linewidth]{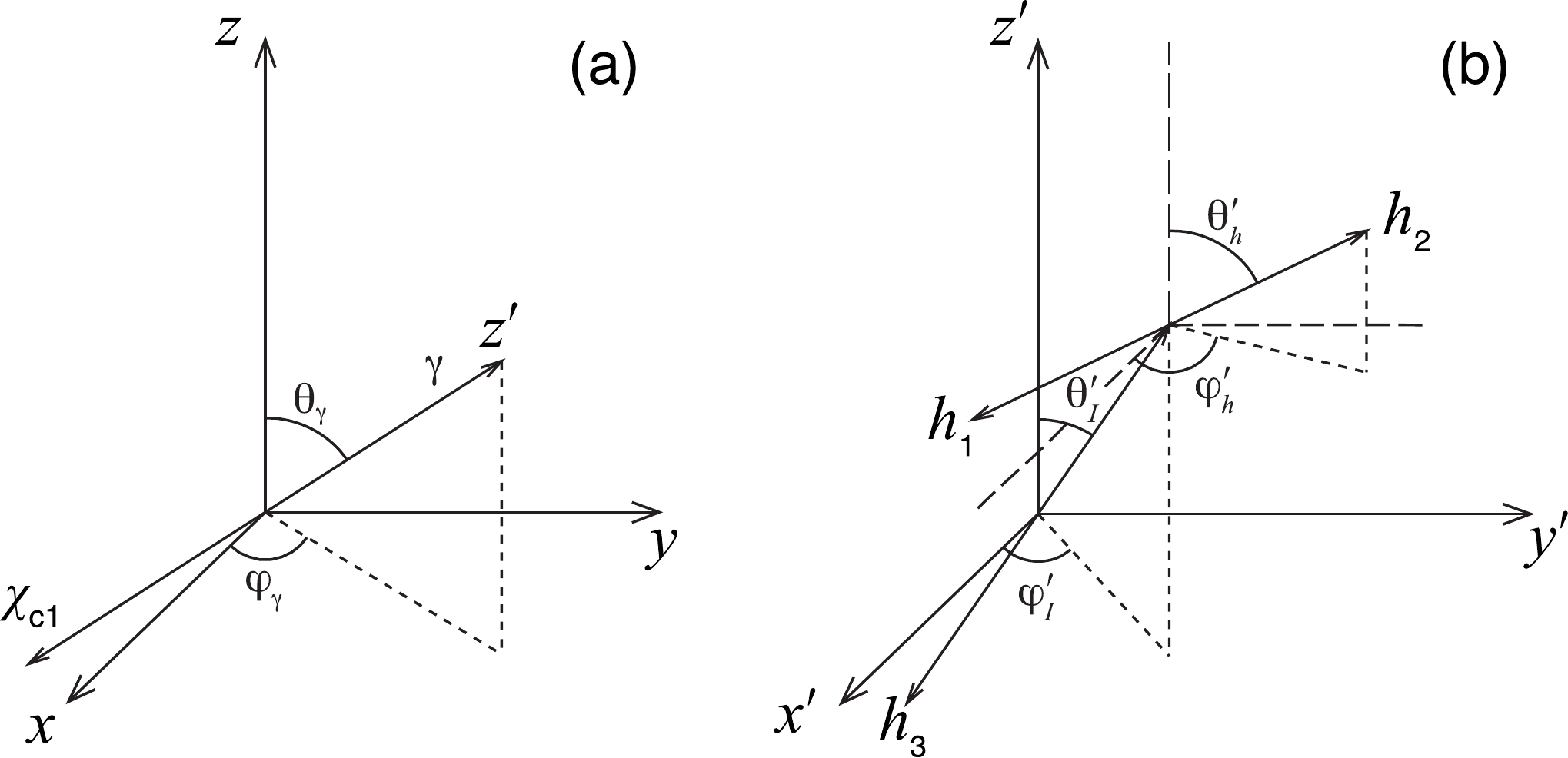}
\caption{ \label{fig:helAng}
The angles used to describe the initial $\psi(2S)$ decay (a) and the subsequent decay of the $\chi_{c1}$ (b)
}
\end{center}
\end{figure*}

In order to describe the angular distribution of the final state particles we measure angles in two coordinate systems which are depicted in Fig.~\ref{fig:helAng} and related in Eq.~(\ref{eq:Amp}) by the $D$-function at the end of the first line.  The angles $\theta_{\gamma}$ and $\phi_{\gamma}$ are the polar and azimuthal angle of the radiated photon in the $\psi(2S)$ rest frame, where $\hat{z}$ is given by the $e^+$ beam direction and $\hat{y}$ is (arbitrarily) defined as upward in the laboratory.  (The amplitude is uniform in $\phi_\gamma$.)  

The two spherical harmonics in the second line of Eq.~(\ref{eq:Amp}) provide a description of the angular distribution for the initial $\chi_{c1}$ decay and the subsequent isobar decay for various values of isobar-hadron orbital angular momentum $L$ and isobar angular momentum $J$.  The angles $\theta^\prime_{I}$ and $\phi^\prime_{I}$ are the polar and azimuthal angles of the isobar in the $\chi_{c1}$-helicity frame, defined as the rest frame of the $\chi_{c1}$ with $z^\prime$-axis along the photon momentum and $y^\prime$-axis perpendicular to the plane formed by the $\psi(2S)$ and photon three-momenta.  Finally, the angles $\theta_h^\prime$ and $\phi_h^\prime$ are the polar and azimuthal angles of $h$, one of the hadrons produced in the isobar decay, after boosting the momentum of $h$ in the $\chi_{c1}$-helicity frame to the isobar rest frame~\cite{Ascoli75}.  All values of $M^\prime_L$ and $M^\prime_J$ are summed with appropriate Clebsch-Gordan coefficients to create the initial $\chi_{c1}$ state with one unit of total angular momentum and $z^\prime$ projection $M^\prime_\chi$.

The ``breakup momentum" in a decay of $1\to2$ particles is given by the momentum of one of the produced particles in the rest frame of the parent.  We denote the breakup momentum of the initial $\chi_{c1}$ decay and the secondary isobar decay by $p$ and $q$, respectively.  Finally, the term $T^\alpha(s)$, described in detail in the next section, is a function of the invariant mass squared of the isobar and describes the two-body dynamics in the decay.

To impose isospin symmetry in the decays $\chi_{c1}\to a_J^\pm\pi^\mp$, we write the $a_J\pi$ amplitude as
\begin{equation}
A^{a_J\pi}_{M_\psi,\lambda_\gamma}(\mathbf{x}) = \frac{1}{\sqrt{2}}\left( A^{a_J^+\pi^-}_{M_\psi,\lambda_\gamma}(\mathbf{x})+A^{a_J^-\pi^+}_{M_\psi,\lambda_\gamma}(\mathbf{x})\right),
\end{equation}
where the distinction between the two terms on the right-hand side is the interchange of the $\pi^+$ and $\pi^-$ four-momenta in the calculation of the relevant kinematic variables.  A similar symmetrization is used in the construction of the $\pi_1\pi$ amplitude.

\subsubsection{Two-body dynamics}

We use three different formulations of $T_\alpha(s)$ [in Eq.~(\ref{eq:Amp})] to describe the isobar decay amplitude and phase as a function of $s$, the invariant mass squared of the isobar decay products.  For all intermediate states except the $a_0(980)$ and the $\pi\pi~S$-wave we use a Breit-Wigner distribution,
\begin{equation}\label{eq:BW}
T_\alpha(s) = \frac{ 1 }{ m^2_0 - s -im_0\Gamma_{J}(s)},
\end{equation}
with
\begin{equation}\label{eq:GammaBW}
\Gamma_{J}(s) = \Gamma_0  \frac{\rho(s)}{\rho_0}  \left [ \frac{B_J\left(q(s)\right)}{B_J(q_0)} \right ]^2,
\end{equation}
where $m_0$ and $\Gamma_0$ are the isobar mass and width, respectively.  We define the breakup momentum $q_0\equiv q(m_0^2)$.  Likewise, the available phase space is given by $\rho(s) = 2q(s)/\sqrt{s}$, and $\rho_0 \equiv \rho(m_0^2)$.  These factors are used in conjunction with $B_J(q)$, a spin-dependent Blatt-Weisskopf barrier penetration factor~\cite{Blatt51}, to construct the mass-dependent total decay width given in Eq.~(\ref{eq:GammaBW}).

To describe the $\aze$ line-shape we use a three-channel Flatt\'e formula~\cite{FLATTE76}.  In addition to the common decay modes
$\aze\rar\eta\pi$ and $\aze\rar KK$, we include a third decay mode, $\aze\rar\eta'\pi$, to provide a consistent description for both the $\chin\rar\eta\pip\pim$ and $\chin\rar\eta'\pip\pim$ data.  The parametrization takes the form 
\begin{equation}\label{eq:Flatte}
T_{a_0(980)}(s) = \frac{ 1  }{ m^2_0 - s -i\sum_c g_c^2\rho_c },
\end{equation}
where $m_0$ is the $\aze$ mass and $g_c$ represents a coupling to one of the $a_0(980)$ decay modes: $\eta\pi$, $KK$, or $\etaprime\pi$.  The factors $\rho_c$ are the phase space available for each of the three different final states.  Following the technique suggested by Flatt\'e to preserve analyticity at the $KK$ and $\eta^\prime\pi$ thresholds, we allow the phase space factors to become imaginary when $s$ is below threshold for a particular decay channel~\cite{FLATTE76}.

For the $\pi\pi$ $S$-wave, we utilize an analysis of $\pi\pi$ scattering data~\cite{LEON9499} that provides two independent amplitudes for $\pi\pi\rar\pi\pi$ and $KK\rar\pi\pi$ production mechanisms.  Specifically, we attempt to model both direct production of $\chin\rar\eta\pi\pi$ with the $\pi\pi$ in an $S$-wave and also the process $\chin\to\eta K K \rar\eta\pi\pi$ where the $K K$ $S$-wave intermediate state rescatters into $\pi\pi$ $S$-wave.  These two amplitudes, labeled $\spp$ and $\skk$ respectively, are constructed to be consistent with existing data in the region of $\pi\pi$ invariant mass below 2~GeV/$c^2$, where the $S$-wave is expected to be significant.

\begin{figure}
\begin{center}
\ifthenelse{\equal{\usepreprint}{1}}{
\includegraphics[width=0.3\linewidth]{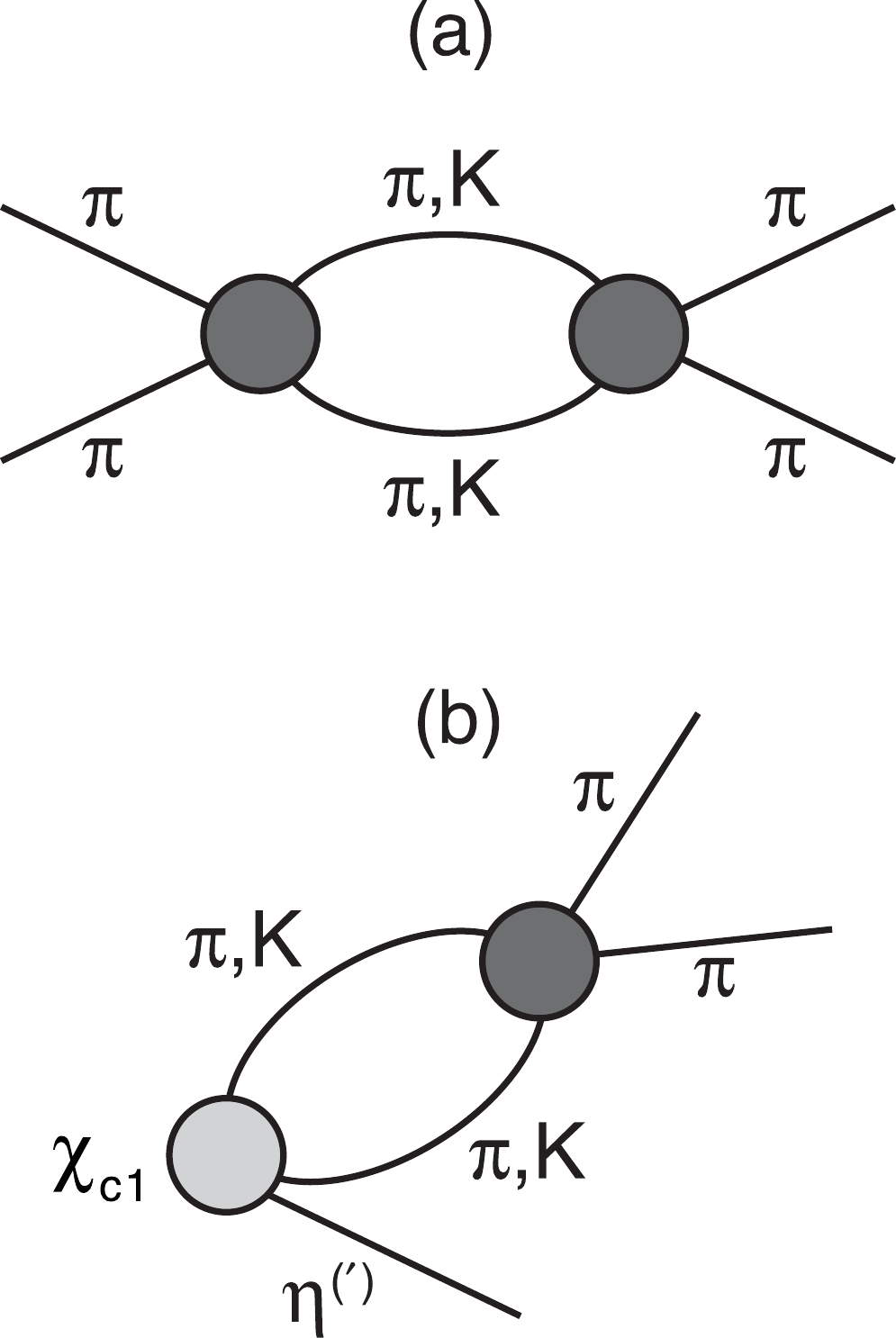}
}{
\includegraphics[width=0.6\linewidth]{4250911-007}
}
\caption{ \label{fig:ppDiagrams}
A diagram of the $\pi\pi$ scattering process (a) and the $\chi_{c1}$ decay process (b).  Both processes may have intermediate pion or kaon loops.  The dark grey interaction represents that obtained from scattering data, while the light grey interaction is some unknown $s$-dependent production amplitude in $\chi_{c1}$ decay.
}
\end{center}
\end{figure}

\begin{figure*}
\begin{center}
\includegraphics[width=0.7\linewidth]{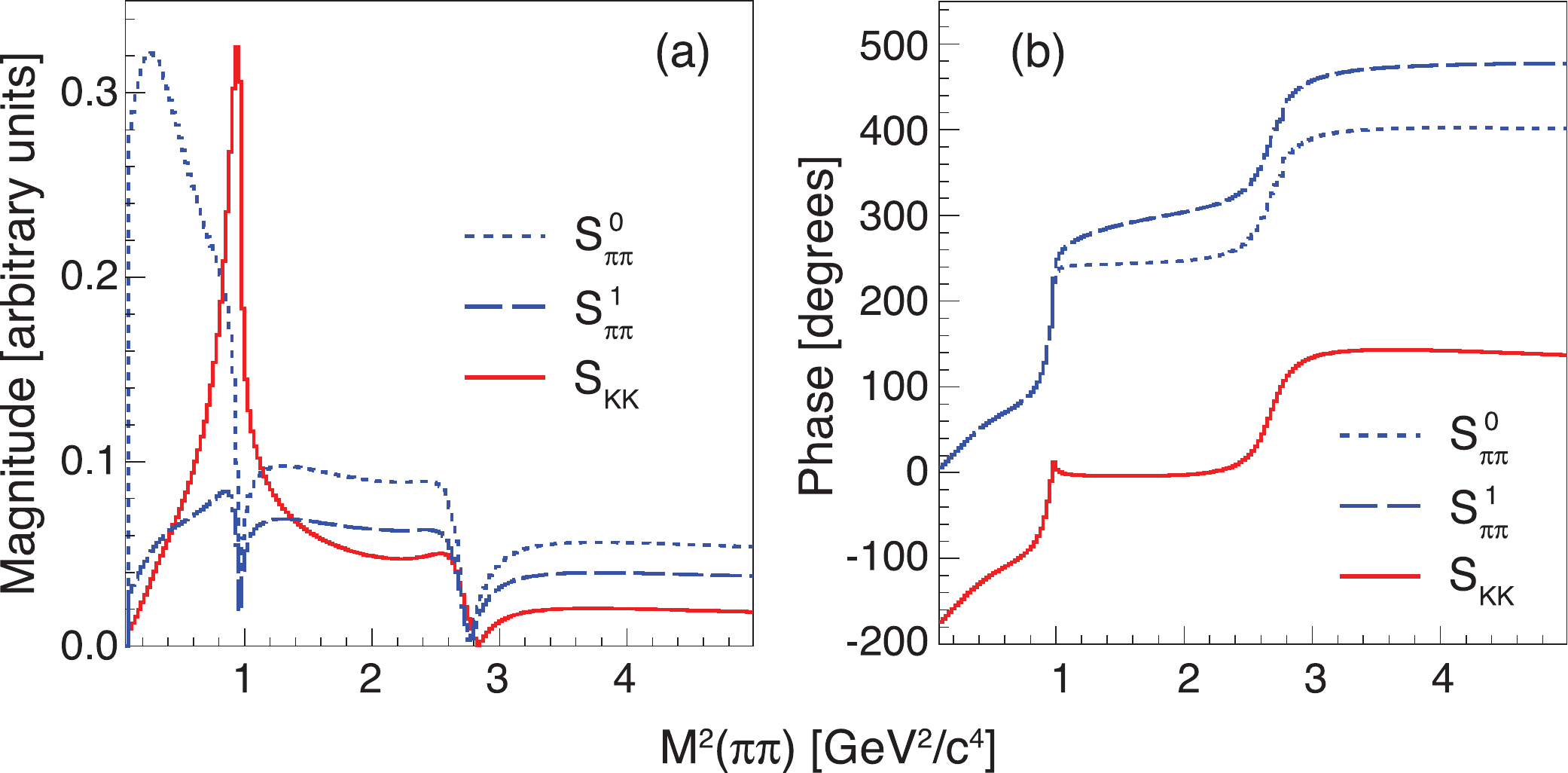}
\caption{ \label{fig:ppSwave}
The (a) magnitudes and (b) phases of both terms of the $\pi\pi\rar\pi\pi$ (dashed and dotted, blue online)
amplitude and the $KK\rar\pi\pi$ (solid, red online) amplitude as a function of the 
two-pion invariant mass squared.}
\end{center}
\end{figure*}

To account for the $s$-dependent differences between $\pi\pi$ scattering, from which the amplitudes are derived, and production in $\chi_{c1}$ decay (see Fig.~\ref{fig:ppDiagrams}), the $\spp$ scattering amplitude is rewritten in a form $N(s)/D(s)$, and the numerator is replaced by the first two terms in a series expansion
\begin{equation}\label{eq:Spp}
S_{\pi\pi}(s) = \frac{1 + z(s)}{D(s)} = \sppz(s) + c~\!\sppo(s)
\end{equation}
where the conformal transform $z(s)$ is given by
\begin{equation}\label{eq:Zsubs}
z(s) = \frac{ \sqrt{s + s_0} - \sqrt{ 4m_K^2 - s } }{ \sqrt{s + s_0} + \sqrt{ 4m_K^2 - s } }.
\end{equation}
The parameter $s_0$, used to adjust the left-hand cut in the complex plane, is set 
to $s_0=1.5$~GeV$^2/c^4$.  Since the $\skk$ amplitude only peaks in a narrow region of $s$, it is assumed that a similar $s$-dependent modification of the production amplitude is not necessary.  The magnitudes and phases of the $\skk$, $\sppz$, and $\sppo$ amplitudes used in the fit are shown in Fig.~\ref{fig:ppSwave}.

Given the definitions above, we can write an expression for the dynamical portion of the $\pi\pi~S$-wave amplitude in the fit:
\begin{equation}
T_{(\pi\pi)_S}(s) = \sppz(s) + c~\!\sppo(s) + k~\!\skk(s),
\label{eq:tpipis}
\end{equation}
where $c$ and $k$ are real parameters that determine the relative sizes of the components.

\subsection{Fitting technique}
\label{sec:fitting}

The extended maximum likelihood method is utilized to determine the best model parameters to describe the data.  The likelihood can be written as
\begin{equation}
\mathcal{L} = \frac{e^{-\mu}\mu^N}{N!}\prod^N_{i=1}\frac{\zeta(\mathbf{x}_i)I(\mathbf{x}_i)}{\int \zeta( \mathbf{x} )I(\mathbf{x})d\mathbf{x}},
\end{equation}
where $\mathbf{x}$, as above, is a position in the multi-dimensional space that spans the event kinematics. The functions $\zeta$ and $I$ describe the efficiency of the analysis criteria and the model-predicted density of events in this space, respectively.  The total number of observed events is $N$, and the model-predicted number of events is represented by
\begin{equation}
\mu \equiv \int\zeta( \mathbf{x} )I(\mathbf{x})d\mathbf{x}.
\end{equation}
In practice one varies the fit parameters to minimize the function
\begin{equation}
-2\ln\mathcal{L} = 2 \left( \int \zeta(\mathbf{x})I(\mathbf{x})d\mathbf{x} - \sum_{i=1}^N\ln I(\mathbf{x}_i)\right) + \kappa,
\label{eq:loglik}
\end{equation}
where $\kappa$ is a constant term that is not included in the minimization procedure.  The integral on the right hand side of the equation is evaluated using MC techniques to determine the average value of $\zeta(\mathbf{x})I(\mathbf{x})$.  Namely, we generate an MC sample of $N_g$ signal events that are distributed uniformly in phase space.  These events are subjected to our selection criteria and yield a sample of $N_a$ accepted events.  For the current study $N_a/N\approx 60$.  From this sample we can compute
\begin{equation}
 \int\zeta(\mathbf{x})I(\mathbf{x})d\mathbf{x} = 
\mathcal{V}\left\langle \zeta(\mathbf{x})I(\mathbf{x}) \right \rangle \approx
 \frac{\mathcal{V}}{N_g}\sum_{i=1}^{N_a}I(\mathbf{x}_i),
\label{eq:mcint}
\end{equation}
where $\mathcal{V}$ is the volume of phase space that spans the event kinematics.

In the context of the model discussed in the previous section, $I(\mathbf{x})$ is as given in Eq.~(\ref{eq:inten}).  Finally, after substituting Eq.~(\ref{eq:mcint}) into Eq.~(\ref{eq:loglik}), making a change of variables $V_{M_\psi,\lambda_\gamma}^\alpha\to V_{M_\psi,\lambda_\gamma}^\alpha / \sqrt{\mathcal{V}}$, and collecting constant terms into $\kappa^\prime$, we can write the  expression that is minimized by the fitter:
\begin{eqnarray}
&~&\hspace{-.5cm}-2\ln\mathcal{L} - \kappa^\prime = \nonumber \\
&~&~~\frac{2}{N_g} \sum_{i=1}^{N_a} \sum_{M_\psi,\lambda_\gamma}\left|\sum_{\alpha}V_{M_\psi,\lambda_\gamma}^\alpha A_{M_\psi,\lambda_\gamma}^\alpha(\mathbf{x}_i)\right|^2 - \nonumber \\
&~&~~2\sum_{i=1}^{N}\ln\left(  \sum_{M_\psi,\lambda_\gamma}\left|\sum_{\alpha}V_{M_\psi,\lambda_\gamma}^\alpha A_{M_\psi,\lambda_\gamma}^\alpha(\mathbf{x}_i)\right|^2 \right).
\end{eqnarray}
Note that $\kappa^\prime$ is a constant that only depends on $\mathcal{V}$ and the number of events $N$; therefore, its actual value is not needed to construct a ratio of likelihoods ($\mathcal{L}$) for two different models that describe the same set of data.

In the fit, we constrain $V_{1,1}^\alpha = V_{1,-1}^\alpha =V_{-1,1}^\alpha=V_{-1,-1}^\alpha$ for each decay amplitude.  Recall that the coefficients $C$ defined in Eq.~(\ref{eq:alpha}) ensure the proper linear combination of $\psi(2S)$ and photon helicities is used to generate the desired $E1$ or $M2$ transition amplitude.

In order to compare the results of our unbinned fit with the data for a given kinematic variable, we take the signal MC sample that passes our event selection criteria and weight each entry in a histogram of the given variable by the function $I(\mathbf{x})$, defined above.  To isolate contributions from various $\chi_{c1}$ decay modes, we can restrict the sum used to compute the weight $I(\mathbf{x})$. 

The results of the fit can be cast as an acceptance-corrected ``fit fraction" for each $\chi_{c1}$ decay amplitude $\alpha$ given by
\begin{equation}
\mathcal{F}_\alpha \equiv \frac{\int  \sum_{M_\psi,\lambda_\gamma}\left|V_{M_\psi,\lambda_\gamma}^\alpha A_{M_\psi,\lambda_\gamma}^\alpha(\mathbf{x})\right|^2 d\mathbf{x}}{ \int  I(\mathbf{x}) d\mathbf{x}}.
\label{eq:fitfrac}
\end{equation}
The integrals of the amplitudes over phase space are evaluated using MC techniques, similar to Eq.~(\ref{eq:mcint}), except the sum is performed over the $N_g$ generated MC events.  Note that due to interference between the decay amplitudes, $A^\alpha_{M_\psi,\lambda_\gamma}(\mathbf{x})$, the sum of the fit fractions, $\sum_\alpha \mathcal{F}^\alpha$, is not constrained to unity.

We compute errors on the fit fractions by propagating the errors on the fitted values of $V_{M_\psi,\lambda_\gamma}^\alpha$ through Eq.~(\ref{eq:fitfrac}).  We stress this point because such errors are not suitable for estimating the statistical significance of amplitudes with relatively small fit fractions.  We obtain the statistical significance of amplitude $\alpha$ by computing the ratio of the likelihood of the null hypothesis ($\mathcal{F}^\alpha\to0$) to the likelihood of our baseline fit, which is derived by retaining known or possible $\eta^{(\prime)}\pi$ and $\pi\pi$ resonances that have a significance greater than one standard deviation.


\section{\label{Results}Results}

\begin{figure*}
\includegraphics[width=0.9\linewidth]{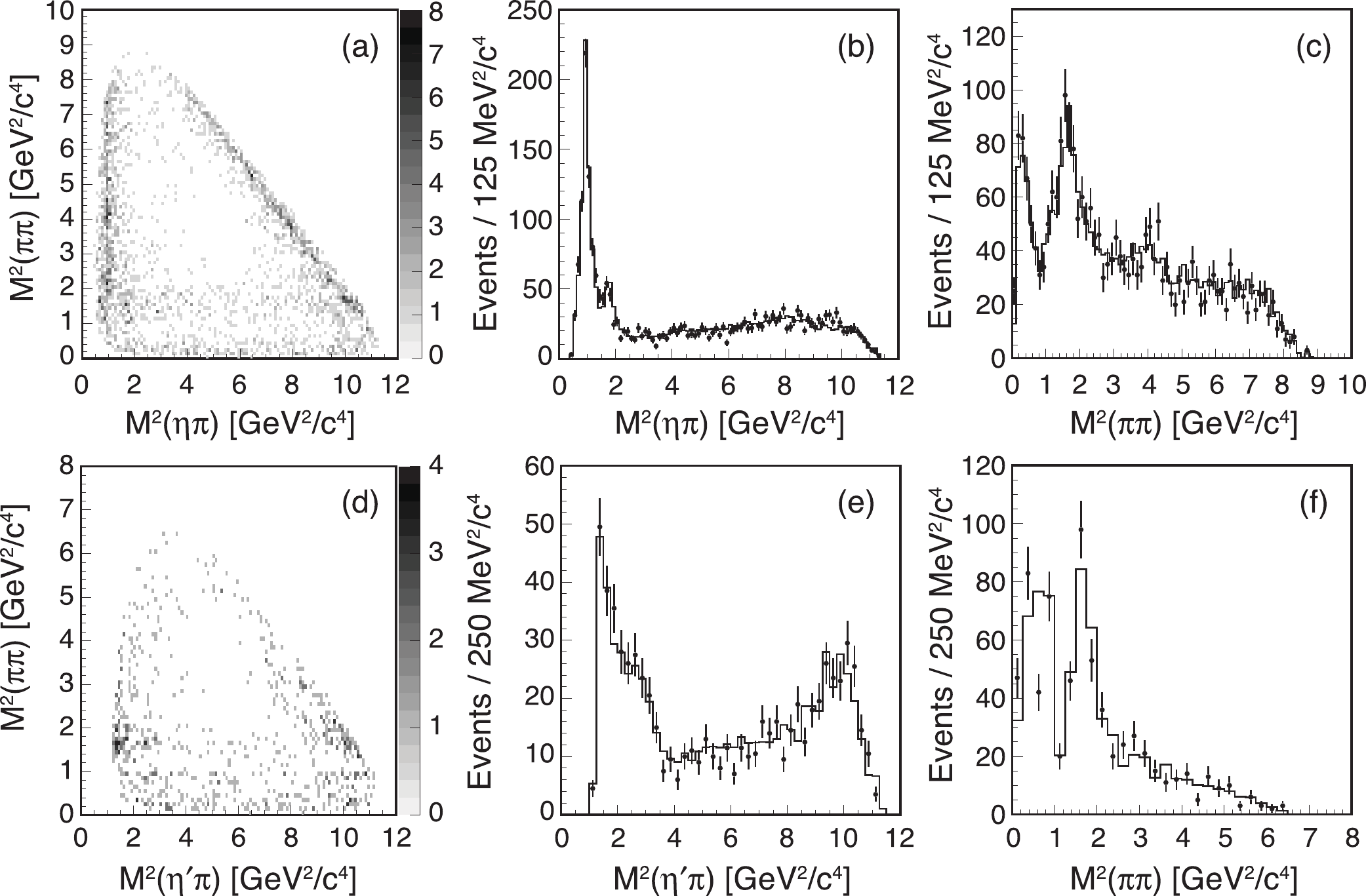}
\caption{Dalitz plots and corresponding projections from the $\chin\rar\eta\pip\pim$ (a-c)
and $\chin\rar\eta'\pip\pim$ (d-f) decays, overlaid with the baseline fits (solid lines).
\label{fig:Dalitz}
}
\end{figure*}

Figures~\ref{fig:Dalitz}(a) and~\ref{fig:Dalitz}(d) display Dalitz plots from the $\chine$ and $\chinep$ decays.  We perform the amplitude analysis, discussed in the previous section, independently on both the $\chine$ and $\chinep$ samples.   In both cases our signal MC sample includes all of the various $\eta$ and $\eta'$ decay modes that we reconstructed in the data populated according to their known branching fractions~\cite{PDG10}.  Figures~\ref{fig:Dalitz}(b) and~\ref{fig:Dalitz}(c), and Figs.~\ref{fig:Dalitz}(e) and~\ref{fig:Dalitz}(f) show projections of the Dalitz plot for the $\chine$ and $\chinep$ data samples, respectively, along with the baseline fit for each sample.

\begin{figure*}
\includegraphics[width=0.85\linewidth]{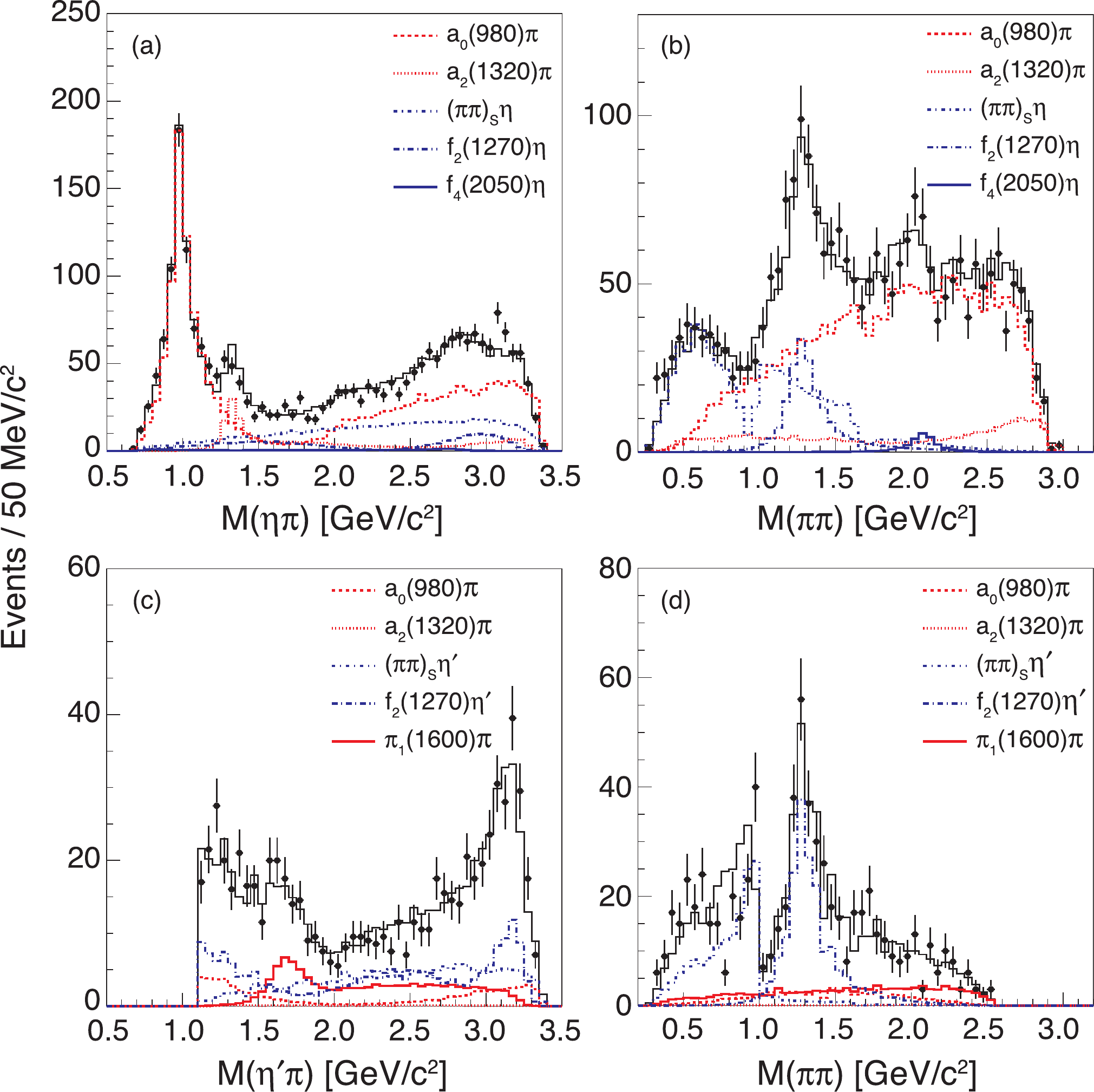}
\caption{Invariant mass projections from the analysis of the $\chin\rar\eta\pip\pim$ (a,b)
 and $\chin\rar\eta'\pip\pim$ (c,d) decays. 
\label{fig:mfits}
}
\end{figure*}

The projections of data and our baseline fits, separated into the various amplitudes, are shown in Fig.~\ref{fig:mfits}.  For those isobars with $J>0$ we try multiple values of $L$ (see Table~\ref{tab:spins}); only in the case of $f_2(1270)\eta^{(\prime)}$ do we find a statistically significant contribution from higher orbital angular momenta.  We report the sum of these $P$-wave and $F$-wave contributions as $f_2(1270)\eta^{(\prime)}$; the individual $P$ and $F$ wave fractions, as expected, have strongly anti-correlated statistical and systematic errors.  We also separate the various $(\pi\pi)_S\eta^{(\prime)}$ contributions given in Eq.~(\ref{eq:tpipis}).  In the fits to the $\chinep$ sample we fix the $\pi\pi~S$-wave parameter $c$ in Eq.~(\ref{eq:tpipis}) to zero; allowing this parameter to float yields a value that is statistically consistent with both zero and the value obtained in the higher-statistics $\chine$ fits.  In all cases the parameters describing the masses and widths of the intermediate resonances are fixed in our baseline fits to enhance the stability of the fit.  Both the $\pi_1(1600)$ and $\ffo$ parameters are fixed to values that maximize the likelihood.  We systematically explore uncertainties on the parametrization of the amplitudes as discussed in Section~\ref{Systematics}.

A quantitative summary of the baseline fits appears in Table~\ref{tab:results}.  From the fit one can compute the total acceptance-corrected event yield in either the $\eta^\prime\pip\pim$ or $\eta\pip\pim$ final states, which is the denominator of Eq.~(\ref{eq:fitfrac}).  If we denote this quantity $N(\eta\pi^+\pi^-)$ or $N(\eta^\prime\pi^+\pi^-)$, respectively, then we can compute the branching fractions for $\chi_{c1}$ to these final states, $\mathcal{B}(\chi_{c1}\to\eta\pi^+\pi^-)$ and $\mathcal{B}(\chi_{c1}\to\eta^\prime\pi^+\pi^-)$, as 
\begin{eqnarray}
&~&\hspace{-.5cm} \mathcal{B}(\chi_{c1}\to\eta^{(\prime)}\pi^+\pi^-) = \nonumber \\
&~&~~~ \frac{p~\!N(\eta^{(\prime)}\pi^+\pi^-)}{N_{\psi(2S)}\mathcal{B}(\psi(2S)\to\gamma\chi_{c1})\sum_i\mathcal{B}_i(\eta^{(\prime)})},
\end{eqnarray}
where $N_{\psi(2S)}$ is the number of initial $\psi(2S)$, $2.59\times10^{7}$, and we use ${\cal B}(\psichin) = (9.2\pm0.4) \times 10^{-2}$~\cite{PDG10}.  The sum over $\eta$ and $\eta^\prime$ branching fractions encompasses all $\eta$ and $\eta^\prime$ decay modes in our signal MC sample, indicated in Table~\ref{tab:fstates}.  The value $p$ is the purity of the data sample in the $\chi_{c1}$ region in Fig.~\ref{fig:mChic1}, which is obtained as discussed in Section~\ref{DataSelect}.

In what follows we discuss the results of the fits to each of the samples in detail, highlighting the key results obtained from each fit.  For each $\chin$ decay mode, we also compute the product $\mathcal{B}(\chi_{c1}\to\eta^{(\prime)}\pi^+\pi^-)\times\mathcal{F}$, which can be interpreted as the branching fraction for the $\chin$ decay to the isobar and spectator multiplied by the branching fraction for the isobar to decay to the $\eta^{(\prime)}\pi^\pm$ or $\pi^+\pi^-$ final state.  Dividing products with common factors (discussed in Section~\ref{sec:ratios})  yields $\chi_{c1}$ and isobar branching ratios.

\begin{table*}
\caption{\label{tab:results}
Summary of results of the baseline fits.  The first and second errors are statistical and systematic, respectively.  The third error, where reported, is from the external value of $\mathcal{B}(\psi(2S)\to\gamma\chi_{c1})$.  Amplitudes that are preceded by an asterisk~(*) are not part of the baseline fits but have been included to determine upper limits.  The listed fit fractions and significances ($N_\sigma$) are obtained when the amplitude is added to the baseline fits.
}
\medskip
\begin{center}
\begin{tabular}{lccc}
\hline
\hline
$\chi_{c1}$ Decay Mode   &   \hspace{1cm} $ {\cal F}$  $[\%]$\hspace{1cm}  & $\mathcal{B}(\chi_{c1}\to\eta^{(\prime)}\pi^+\pi^-)\times\mathcal{F}$ [$10^{-3}$] & \hspace{1cm} $N_{\sigma}$\hspace{1cm} \\
\hline
$\eta\pip\pim$ 	&	-							&	$	4.97	\pm	0.08	\pm	0.21	\pm	0.22	$	&	-		\\																								
\hspace{0.20cm} $\aze\pi$	&	$	66.2	\pm	1.2	\pm	1.1	$	&	$	3.29	\pm	0.09	\pm	0.14	\pm	0.15	$	&	 $>10$		\\	
\hspace{0.20cm} $\atw\pi$	&	$	9.8	\pm	0.8	\pm	1.0	$	&	$	0.49	\pm	0.04	\pm	0.05	\pm	0.02	$	&	9.7		\\	
\hspace{0.20cm} $(\pip\pim)_{S}\eta$	&	$	22.5	\pm	1.3	\pm	2.5	$	&	$	1.12	\pm	0.06	\pm	0.13	\pm	0.05	$	&	$>10$		\\	
\hspace{0.50cm} $\sppz\eta$ 	&	$	12.1	\pm	1.7	\pm	5.6	$	&	$	0.60	\pm	0.08	\pm	0.28	\pm	0.03	$	&	 $>10$		\\	
\hspace{0.50cm} $\sppo\eta$	&	$	3.4	\pm	0.9	\pm	1.5	$	&	$	0.17	\pm	0.05	\pm	0.07	\pm	0.01	$	&	 6.0		\\	
\hspace{0.50cm} $\skk\eta$	&	$	3.1	\pm	0.6	\pm	0.4	$	&	$	0.15	\pm	0.03	\pm	0.02	\pm	0.01	$	&	9.4		\\	
\hspace{0.20cm} $\ftw\eta$	&	$	7.4	\pm	0.8	\pm	0.6	$	&	$	0.37	\pm	0.04	\pm	0.04	\pm	0.02	$	&	 $>10$		\\	
\hspace{0.20cm} $\ffo\eta$	&	$	1.0	\pm	0.3	\pm	0.3	$	&	$	0.05	\pm	0.01	\pm	0.02	\pm	0.00	$	&	5.2		\\	
\hspace{0.20cm} \!\!\!*$\pione\pi$	&		-						&	$<	0.031	$							&	0.7		\\	
\hline																							
$\etaprime\pip\pim$	&	-							&	$	1.90	\pm	0.07	\pm	0.08	\pm	0.09	$	&	-		\\																								
\hspace{0.20cm} $\aze\pi$	&	$	11.0	\pm	2.3	\pm	1.8	$	&	$	0.21	\pm	0.04	\pm	0.04	\pm	0.01	$	&	 8.4		\\	
\hspace{0.20cm} $\atw\pi$ 	&	$	0.4	\pm	0.5	\pm	0.6	$	&	$<	0.031	$							&	1.4		\\	
\hspace{0.20cm} $(\pip\pim)_{S}\eta$	&	$	21.6	\pm	2.7	\pm	1.2	$	&	$	0.41	\pm	0.05	\pm	0.03	\pm	0.02	$	&	10.2		\\	
\hspace{0.50cm} $\sppz\etaprime$ 	&	$	7.0	\pm	2.2	\pm	2.3	$	&	$	0.13	\pm	0.04	\pm	0.04	\pm	0.01	$	&	 6.6		\\	
\hspace{0.50cm} $\skk\etaprime$	&	$	8.4	\pm	1.5	\pm	1.3	$	&	$	0.16	\pm	0.03	\pm	0.02	\pm	0.01	$	&	 7.5		\\	
\hspace{0.20cm} $\ftw\etaprime$ 	&	$	27.0	\pm	2.9	\pm	1.7	$	&	$	0.51	\pm	0.06	\pm	0.04	\pm	0.03	$	&	$>10$		\\	
\hspace{0.20cm} \!\!\!*$\ffo\etaprime$	&		-						&	$<	0.010	$							&	0.4		\\	
\hspace{0.20cm} $\pione\pi$	&	$	15.1	\pm	2.7	\pm	3.2	$	&	$	0.29	\pm	0.05	\pm	0.06	\pm	0.01	$	&	7.2		\\	
\hline\hline

\end{tabular}
\end{center}
\end{table*}

\subsection{\boldmath $\decaya$ decays}\label{EtaPiPi} 

The $\eta\pi^\pm$  and $\pip\pim$ invariant mass projections and corresponding amplitude contributions from the fit to the $\fulldecaya$ sample are shown in Fig.~\ref{fig:mfits}(a,b).  The dominant amplitude in this data set is the $\aze$, which, consequently, must be adequately parametrized to obtain a satisfactory fit to the data.  To determine the $\aze$ parameters, we exclude the data with $\pip\pim$ invariant mass below 1.7~$\gevcc$, which removes any correlation with the $\pip\pim$ $S$-wave amplitudes. 
The fit to this restricted data set includes the $\aze$, $\atw$ and $\ffo$ amplitudes, and we allow all four $\aze$ parameters to float.
The resulting $\aze$ parameters are given in Table~\ref{tab:xlow_a0par}, where the first error is statistical and the second error 
is systematic, obtained by trying various combinations of $\pi\pi$ isobars 
to fit the region in $\pip\pim$ invariant mass around 2.0~$\gevcc$, a peak attributed to the $\ffo$ resonance in the baseline fit.
The $\aze$ Flatt\'e distribution parameters, which are consistent with a previous determination by CLEO~\cite{CLEOCHI}, are subsequently fixed in the baseline fit to the full data sample.  It is worth noting that the $\aze$ lineshape in $\eta\pi$ is rather insensitive to the $\aze\to\eta^\prime\pi$ coupling $g_{\eta^\prime\pi}$.  In fact, the fit prefers a coupling of zero, but with large uncertainty.  Our analysis of $\chi_{c1}\to\eta^\prime\pi^+\pi^-$ data, presented later, directly extracts information related to this coupling constant.

\begin{table}
\caption{\label{tab:xlow_a0par}
The values of $a_0(980)$ parameters compared to the previous CLEO analysis~\cite{CLEOCHI}.
The first error is statistical and the second error is systematic, as explained in the text.
}
\medskip
\begin{center}
\begin{tabular}{ccc}
\hline \hline
Parameter &   [$\gevcc$]    &  Ref.~\cite{CLEOCHI}  [$\gevcc$]  \\
\hline
 $m_0$         & $0.998\pm0.006\pm0.015$ & $1.002\pm0.018$ \\
 $g_{\eta\pi}$ & $0.60\pm0.02\pm0.03$ & $0.64\pm0.05$ \\
 $g_{KK}$      & $0.56\pm0.06\pm0.09$ & $0.52\pm0.15$ \\
 $g_{\eta'\pi}$ & $0.00\pm0.15\pm0.07$ & - \\ 
\hline \hline
\end{tabular}
\end{center}
\end{table}

The $\pi\pi$ $S$-wave is parametrized as described in Eq.~(\ref{eq:tpipis}) with the parameters $c$ and $k$ floating in the fit.  In Table~\ref{tab:results} we list the contributions of the three individual components of the $\pi\pi$ $S$-wave.  In principle, the magnitude and phase of the total $\pi\pi$ $S$-wave can be constructed by using the entries in this table to normalize three components depicted in Fig.~\ref{fig:ppSwave}.

In order to fit the $\pip\pim$ invariant mass distribution around 2.0~$\gevcc$,
we tried various known $\pi\pi$ resonances with $J=0,2,$~and~4 and masses ranging  from 2.0 to 2.3~$\gevcc$.
The best fit is obtained with a single spin-four resonance that has parameters consistent with the $\ffo$ state listed by the PDG~\cite{PDG10}.
The mass and width of the $\ffo$, as determined by our fit, are 
 $m_0 = 2.080\pm 0.025 \pm 0.010$~$\gevcc$ and $\Gamma = 0.160\pm 0.035\pm 0.040$~$\gevcc$.
The systematic errors are obtained by varying the $\aze$ parameters
within the errors listed in Table~\ref{tab:xlow_a0par}.

\subsection{\boldmath $\decayb$ decays}\label{EtaprimePiPi} 

The set of amplitudes used to fit the $\chin\rar\eta'\pip\pim$ data is
listed in Table~\ref{tab:results}, and the corresponding fit projections are shown in Fig.~\ref{fig:mfits}(c,d).
The dominant isobar in this $\chin$ decay mode is the $\ftw$, coupled with the $\eta^\prime$ in both $P$ and $F$ waves.
Two key results emerge from the analysis of the $\eta^\prime\pi$ spectrum.  First, a $P$-wave intensity is necessary to describe the $\eta^\prime\pi^\pm$ mass spectrum in the region of $1.7~\gevcc$.  Second, we find that $\aze\to\etaprime\pi$ decays populate the $\etaprime\pi^{\pm}$ region near threshold.  We discuss both of these findings in the subsections below.

\subsubsection{Evidence of the $P$-wave $\eta'\pi$ amplitude}\label{P1sig} 

\begin{figure*}
\includegraphics*[width=0.9\linewidth]{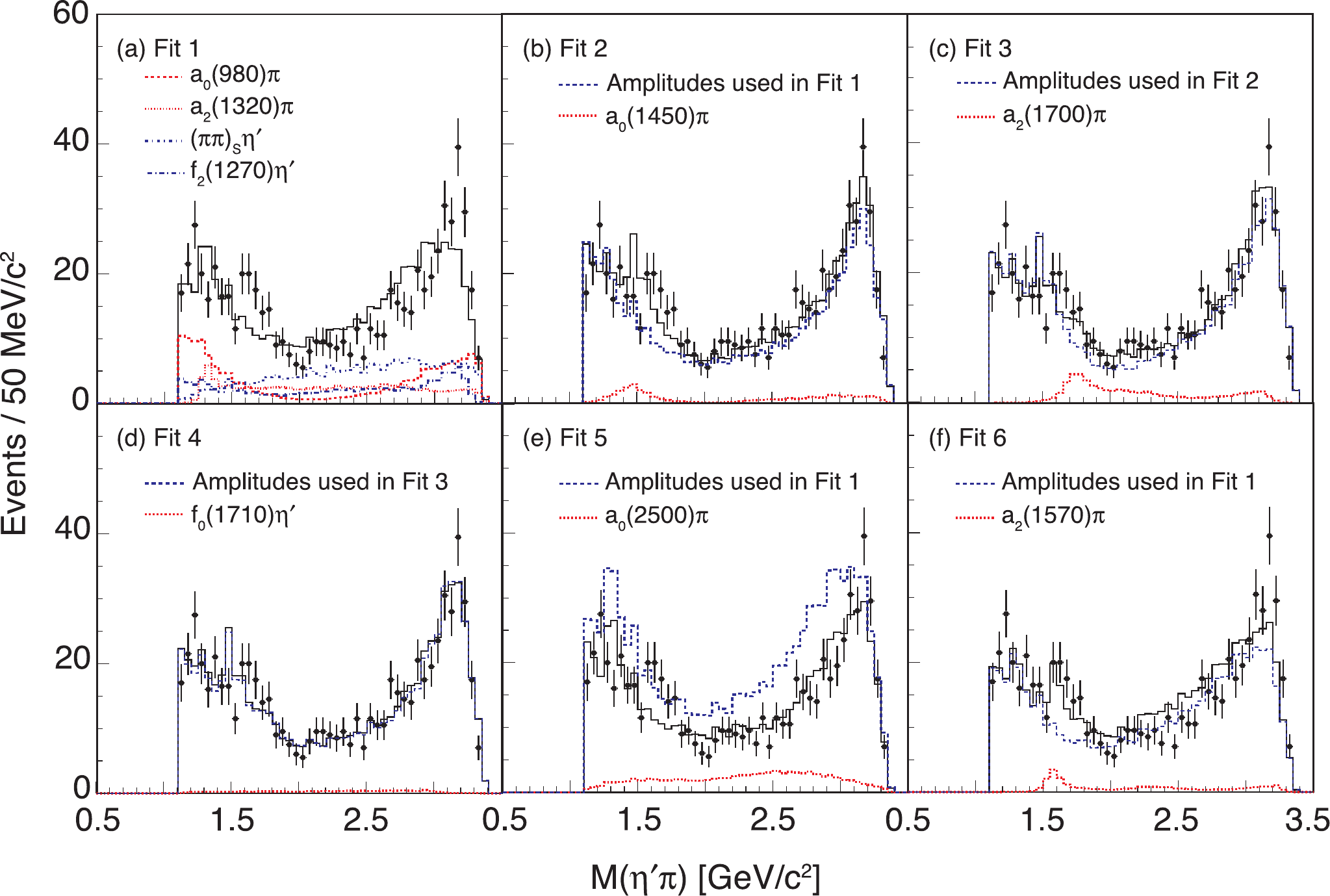}
\caption{Projections of the $\eta'\pi$ invariant mass for fits to the $\chinep$ data when 
the exotic $\pi_1(1600)$ amplitude is excluded. Fits are numbered according to Table~\ref{tab:pi1_alt_fits}.  In all figures the sum of all amplitudes in the fit is indicated by the solid black histogram.
\label{fig:epPP_xP1}
}
\end{figure*}

A fit to the data without the $\chi_{c1}\to\pi_1\pi$ amplitude is shown in Fig.~\ref{fig:epPP_xP1}(a).  It poorly describes the data in in the $\eta'\pi$ invariant mass region near $1.7~\gevcc$ and greater than $2.3~\gevcc$ (due to the contribution from the isospin-conjugate channel).  Our baseline fit accounts for these deficiencies by introducing a $\chi_{c1}\to\pi_1\pi$ amplitude where the $\pi_1$ resonance shape is described by Eq.~(\ref{eq:BW}) and the parameters for the mass and width are determined by the fit.  Such a resonance has exotic quantum numbers, $J^{PC}=1^{-+}$, and cannot be a $q\bar{q}$ meson.

The statistical significance of the $\chi_{c1}\to\pi_1\pi$ amplitude can be evaluated by examining the ratio of maximum likelihoods for the two fits.  If we define
\begin{equation}
\Lambda\equiv \frac{\mathcal{L}(\pi_1\pi~\mathrm{excluded})}{\mathcal{L}(\pi_1\pi~\mathrm{included})},
\end{equation}
then, in the absence of any true $\pi_1\pi$ amplitude, $-2\ln\Lambda$ will approach, in the limit of infinite statistics, a $\chi^2$ distribution with the number of degrees of freedom equal to the number of additional free parameters in the fit that includes the $\pi_1\pi$ amplitude.  As indicated in the first line of Table~\ref{tab:pi1_alt_fits}, the value of $-2\ln\Lambda$ for the baseline fit with and without the $\pi_1\pi$ amplitude is 53, and three additional free parameters are used to describe the $\pi_1\pi$ amplitude.  This results in a very small probability ($\approx 10^{-11}$) that data are a fluctuation of the model used in the fit to Fig.~\ref{fig:epPP_xP1}(a), which does not have a $\pi_1\pi$ amplitude.

In order to search for and quantify the significance of other non-exotic alternatives to the $\pi_1\pi$ amplitude that is used in our baseline fit, we considered several alternative fits to the data, enumerated 2-6 in Table~\ref{tab:pi1_alt_fits} and described briefly here.  Fit 2 replaces the exotic amplitude with the $a_0(1450)$, a state that has a known $\eta'\pi$ decay channel.  Fit 3 adds an $a_2(1700)$ amplitude, which might also decay into the $\eta'\pi$ 
final state.  Fit 4 adds an $f_0(1710)$ state in the $\pi\pi$ channel; heavier $\pi\pi$ states of various spins produce no significant improvements in fit quality.  Finally, fits 5 and 6 test the $J=1$ assignment of the $\eta'\pi$ state by attempting to fit the data with a new $a_0$ or $a_2$ $\eta^\prime\pi$ state whose mass and width are floating in the fit.  The fit returns a value of the mass and width of $2.5~\gevcc$ and $1.4~\gev$ ($1.6~\gevcc$ and $0.1~\gev$) for the new $a_0$ ($a_2$) state, respectively.  In the third column of Table~\ref{tab:pi1_alt_fits} we list the change in $-2\ln\mathcal{L}$ from fit 1, the baseline fit that does not include an exotic amplitude.  None of the alternate fits produces a change in likelihood as significant as the fit that includes the $\pi_1\pi$ amplitude.  Furthermore, we can test the significance of the $\pi_1\pi$ amplitude in the presence of these alternate models by including this amplitude in each of the alternate fits.  The resulting values for $-2\ln\Lambda$ are listed in Table~\ref{tab:pi1_alt_fits}.  Including the $\pi_1\pi$ amplitude introduces just one extra degree of freedom since the mass and width of the $\pi_1$ are fixed to the values obtained in the baseline fit.  The $\pi_1\pi$ amplitude is least significant in the context of fit 5, but even in this fit the significance of the $\pi_1\pi$ amplitude is 4.7 ($\sqrt{22}$) standard deviations.

\ifthenelse{\equal{\usepreprint}{1}}{
\begin{sidewaystable}
}{
\begin{table*}
}
\caption{\label{tab:pi1_alt_fits}
Table of alternate $\pi_1$ fits.  The difference in likelihood (cast as $-2\ln(\mathcal{L}_1/\mathcal{L}_i)$, where $i$ is the fit index) and number of additional free parameters ($N^\mathrm{par}_i - N^\mathrm{par}_1$ ) for each fit when compared to fit 1 is listed on the left side of the vertical line.  The values on the right side of the vertical line are $-2\ln\Lambda$ (see text for definition) and the number of additional free parameters when the $\pi_1\pi$ amplitude is included ($\Delta N^\mathrm{par})$.  The baseline fit corresponds to fit 1 with the $\pi_1\pi$ amplitude included.}
\medskip
\begin{center}
\begin{tabular}{clcc|cc}
\hline\hline
~&~& \multicolumn{2}{c|}{$\pi_1\pi$ amplitude excluded} & \multicolumn{2}{c}{~~$\pi_1\pi$ amplitude included~~} \\
Fit & Description & ~~$-2\ln(\mathcal{L}_1/\mathcal{L}_i)$~~ & ~~$N^\mathrm{par}_{i}-N^\mathrm{par}_1$ ~~& ~~$-2\ln\Lambda$ ~~& $\Delta N^\mathrm{par}$ \\ \hline
1 & Baseline fit without $\pi_1$ & -- & -- & 53 & 3 \\
2 & Fit 1 amplitudes and $a_0(1450)$ & 8 & 1 & 47 & 1 \\
3 & Fit 2 amplitudes and $a_2(1700)$ & 22 & 2 & 34 & 1 \\
4 & Fit 3 amplitudes and $f_0(1710)$ & 30 & 3 & 27 & 1 \\
5 & Fit 1 amplitudes and $a_0$ with floating $m_0$, $\Gamma$ & 34 & 3 & 22 & 1 \\
6 & Fit 1 amplitudes and $a_2$ with floating $m_0$, $\Gamma$ & 32 & 3 & 23 & 1 \\
7 & Fit 1 amplitudes and non-resonant $\eta'\pi$ $P$-wave & 45 & 4 & 10 & 1 \\
\hline\hline
\end{tabular}
\end{center}
\ifthenelse{\equal{\usepreprint}{1}}{
\end{sidewaystable}
}{
\end{table*}
}

It has been noted that intensity in the $P$-wave $\eta^\prime\pi$ scattering amplitude does not have a unique interpretation in terms of QCD and/or hadronic degrees of freedom~\cite{Szczepaniak:2003vg}.  A complete analysis of the amplitude and phase of the $P$-wave is needed.  Unfortunately, it is impossible to extract this information from the data in a model independent way due to the relatively low statistics and three-body nature of this analysis.  Furthermore, non-resonant $\eta^\prime\pi$ $P$-wave interactions are not well constrained, which makes it difficult to systematically test the significance of an exotic resonance in the presence of a non-resonant $P$-wave background.  In an attempt to probe the significance of the Breit-Wigner phase motion, we replace the $\pi_1$ Breit-Wigner parametrization in our baseline fit with an amplitude whose magnitude matches that of a Breit-Wigner function but whose phase is constant (independent of $s$) and a free parameter in the fit.  In this fit, fit 7 in Table~\ref{tab:pi1_alt_fits}, we also float the mass and width of the Breit-Wigner shape that describes the magnitude of the amplitude.  We obtain a mass and width consistent with those obtained by introducing a resonant $\pi_1\pi$ amplitude and the improvement in the fit is 45 units for four additional parameters, which is not as dramatic as is obtained by including a $\pi_1\pi$ amplitude.  The significance of the $\pi_1\pi$ amplitude in the presence of this non-resonant $P$-wave drops to 3 standard deviations.  Our choice of parametrization for non-resonant interactions in this test is somewhat arbitrary -- other choices may yield variations in the significance of the resonant $\pi_1\pi$ amplitude.  While this is suggestive of significant resonant behavior in the $P$-wave, such a definitive conclusion is difficult to make without a complete understanding of non-resonant $P$-wave interactions.  We can, however, state that there is clear evidence for $P$-wave $\eta^\prime\pi$ interactions.

\begin{figure*}
\includegraphics[width=0.7\linewidth]{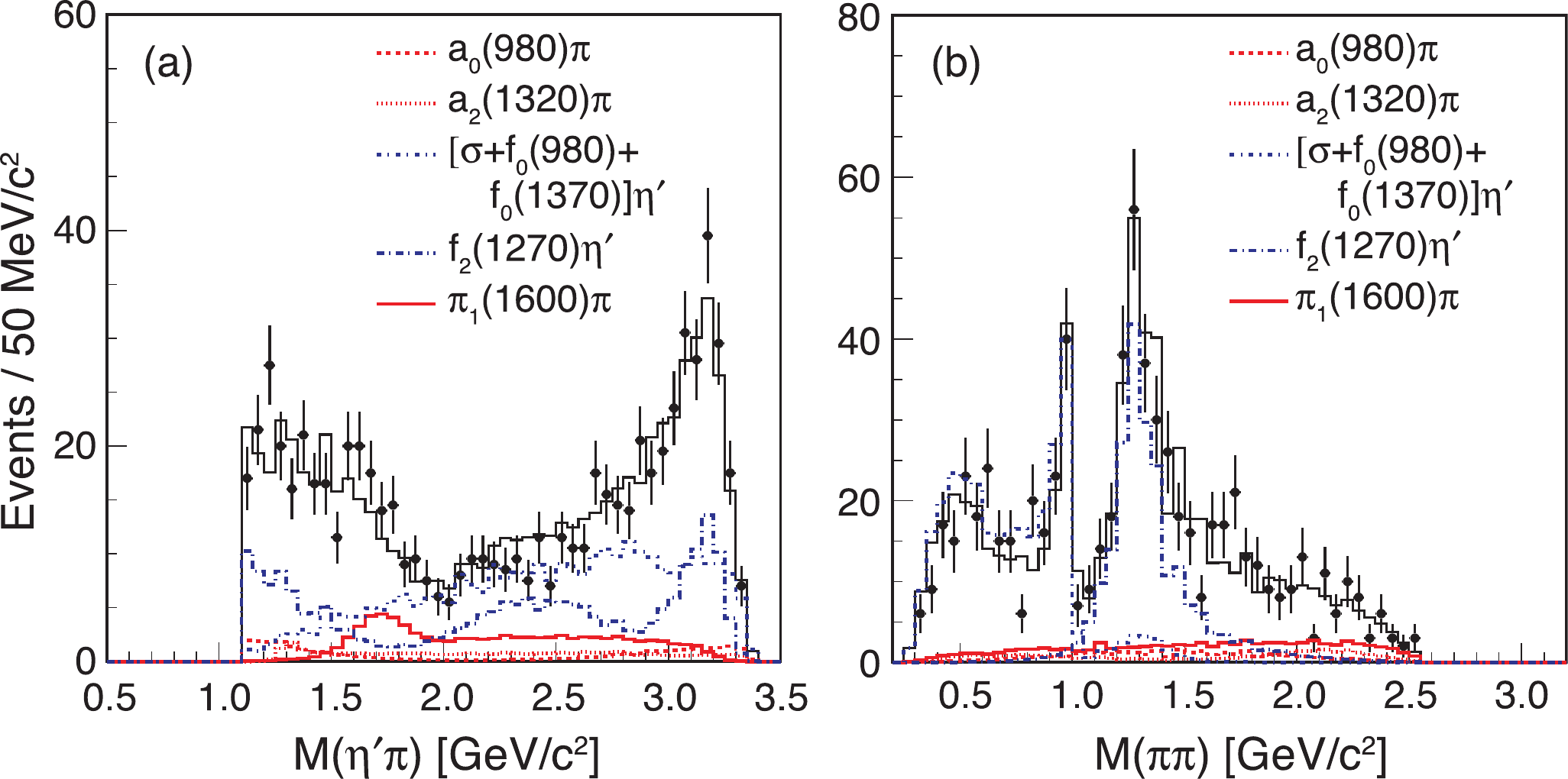}
\caption{\label{fig:altpipi}Invariant mass projections of a fit to the $\eta'\pi\pi$ data that uses an alternate parametrization of the $\pi\pi$ $S$-wave.  (See text for description).}
\end{figure*}

Finally, we examine possible correlations of the $P$-wave intensity with the parametrization of the $\pi\pi$ $S$-wave.  An examination of Fig.~\ref{fig:mfits}(d) suggests that the region of $\pi\pi$ invariant mass below $1.0~\gevcc$ may not be well-described in the baseline fit.  A natural concern is that the significant $\pi_1\pi$ signal might be correlated with the $\pi\pi$ $S$-wave amplitude in this region.  To address this concern we perform two additional sets of fits.  First, we exclude all data with $\pi\pi$ invariant mass below $1.0~\gevcc$, remove the $\pi\pi$ $S$-wave amplitude from our fits, and repeat the exercise outlined above to compare various alternate solutions without a $\pi_1\pi$ amplitude.  We find that including the $\pi_1\pi$ still produces the most significant improvement in the likelihood.  Second, we devise an alternate parametrization of the $\pi\pi$ $S$-wave that is not derived from $\pi\pi$ scattering data.  This parametrization utilizes a complex pole to describe the broad $\pi\pi$ peak often called the $\sigma$ in addition to Breit-Wigner resonances for the $f_0(980)$ and $f_0(1370)$, with parameters set to those provided by PDG averages~\cite{PDG10}.  Such a technique of adding many resonances has been previously used to describe $\pi\pi$ $S$-wave interactions~\cite{pipiexamp}.  The fit with this alternate $\pi\pi$ $S$-wave parametrization is shown in Fig.~\ref{fig:altpipi}, and the fit quality is improved in the region of $\pi\pi$ mass below 1~GeV$/c^2$.  We then repeat fits 1 through 6 listed in Table~\ref{tab:pi1_alt_fits} with this alternate $\pi\pi$ $S$-wave substituted into our baseline fit; the results are qualitatively the same as those derived from our baseline analysis.  The corresponding value of $-2\ln\Lambda$ for fit 1 is 56 units, and the least significant signal for the $\pi_1\pi$ amplitude is in the context of fit 4, where the significance is 4.4 standard deviations.

To summarize, the best fit to the $\chi_{c1}\to\eta'\pi\pi$ data is obtained when a $\pi_1\pi$ amplitude is included, where the $\pi_1$ is described by a Breit-Wigner lineshape with a mass and width of $1670\pm30\pm20~\mevcc$ and $240\pm50\pm60~\mevcc$, respectively.  The significance of a $P$-wave $\eta^\prime\pi$ amplitude is greater than 4 standard deviations under all attempted variations of the fit, some of which are rather extreme and assume the existence of new conventional $a_0$ and $a_2$ states.  While our baseline fit assumes that the $1^{-+}$ $\eta'\pi$ amplitude can be described by a Breit-Wigner resonance, we cannot exclude other non-resonant $P$-wave $\eta'\pi$ interactions that may mimic a $\pi_1$ resonance.  Therefore, we conclude that evidence exists for a $P$-wave $\eta^\prime\pi$ scattering amplitude, which, if parametrized by a single Breit-Wigner resonance, has a mass and width consistent with the $\pi_1(1600)$ reported in other production mechanisms.

Motivated by reports of a $\pi_1(1400)$ observed in the $\eta \pi$ spectrum~\cite{GAMS88,KEK93,E852epp,CBAR98}, we test the significance of an additional $P$-wave $\eta\pi$ and $\eta'\pi$ resonance that has the mass and width of the $\pi_1(1400)$ reported in Ref.~\cite{PDG10}.  In neither the $\eta\pi\pi$ fit nor the $\eta'\pi\pi$ fit is the signal for such a state robust under variations analogous to those in Table~\ref{tab:pi1_alt_fits}, which were used to test the significance of the $\pi_1(1600)$ amplitude.  Averaging over the two charge conjugate decay modes, we obtain the following 90\% confidence level upper limits:  $\mathcal{B}( \chi_{c1} \to \pi_1(1400)^\pm \pi^\mp ) \times \mathcal{B}( \pi_1(1400)^\pm \to \eta \pi^\pm ) < 0.08 \times 10^{-3}$ and $\mathcal{B}( \chi_{c1} \to \pi_1(1400)^\pm \pi^\mp ) \times \mathcal{B}( \pi_1(1400)^\pm \to \eta' \pi^\pm )  < 0.02 \times 10^{-3}$.

\subsubsection{Observation of $\aze\rar\eta'\pi$ decays}\label{a0sig}

\begin{figure}
\ifthenelse{\equal{\usepreprint}{1}}{
\includegraphics[width=0.4\linewidth]{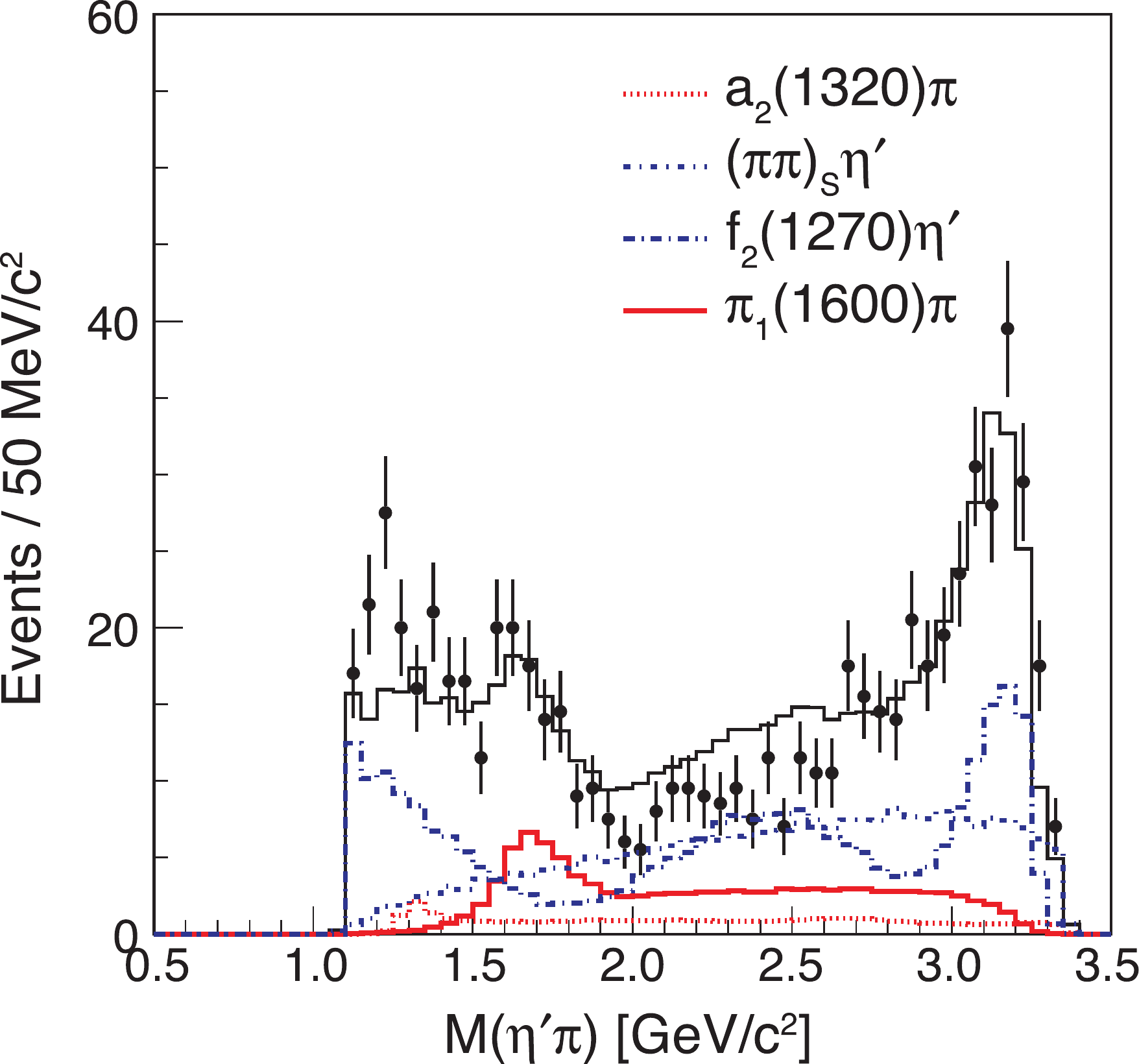}
}{
\includegraphics[width=0.8\linewidth]{4250911-009}
}
\caption{Projection of the $\eta'\pi$ invariant mass for a fit to the $\chinep$ data that excludes the $\aze$ amplitude.  The value of $-2\ln\mathcal{L}$ for this fit, with one fewer free parameter, is 70 units larger than the baseline fit.
\label{fig:epPP_xa0}
}
\end{figure}

Figure~\ref{fig:epPP_xa0} shows a projection of the $\eta'\pi$ invariant mass for a fit that includes all amplitudes in the baseline fit except the $a_0(980)\pi$ amplitude.  This fit does a poor job of describing the data near the $\eta^\prime\pi$ threshold.  Our baseline fit corrects this by introducing an $\aze\pi$ amplitude, where the $\aze$ lineshape utilizes the same Flatt\'e parameters for the mass and couplings to $\eta\pi$ and $KK$ channels obtained from the analysis of the $\eta\pip\pim$ data.  The significance of the $a_0(980)\pi$ amplitude, when compared with this alternate fit, is  8.4 standard deviations.

Like the studies above for the $\pi_1\pi$ amplitude, we try a variety of alternate fits that include the $a_0(1450)$ and $a_2(1700)$ resonances in the $\eta'\pi$ channel and the $f_0(1710)$ and $f_2(1750)$ resonances in the $\pi\pi$ channel to describe the intensity at the $\eta'\pi$ mass threshold.  The maximum change in $-2\ln\mathcal{L}$ observed for any alternate fit was 16 units, compared with 70 units when the $a_0(980)\pi$ amplitude is included.  We can repeat this study with the $\pi_1\pi$ amplitude excluded from the fit and similar results are obtained.  Finally, we also study the stability of the $a_0(980)$ amplitude when the alternate resonance-based parametrization of the $\pi\pi~S$-wave, discussed above, is used.  We find that the overall fit fraction of $a_0(980)$ decreases to about 50\% of the value obtained in our baseline fit; however, the signal remains significant at the level of 6 standard deviations.

\subsection{\label{sec:ratios}Branching ratios}

Using the product branching fractions for the two final states, we can construct two different types of branching ratios.  If we divide products that contain a common $\pi\pi$ isobar, then we can compare the production of this isobar in $\chi_{c1}$ decay against an $\eta$ and $\eta^\prime$.  If we divide products that contain a common $\eta^{(\prime)}\pi$ isobar, the result is measurement of the ratio of $\eta\pi$ to $\eta^\prime\pi$ branching fractions for that isobar.  We discuss each of these types of branching ratios below.

\begin{figure*}
\includegraphics[width=0.7\linewidth]{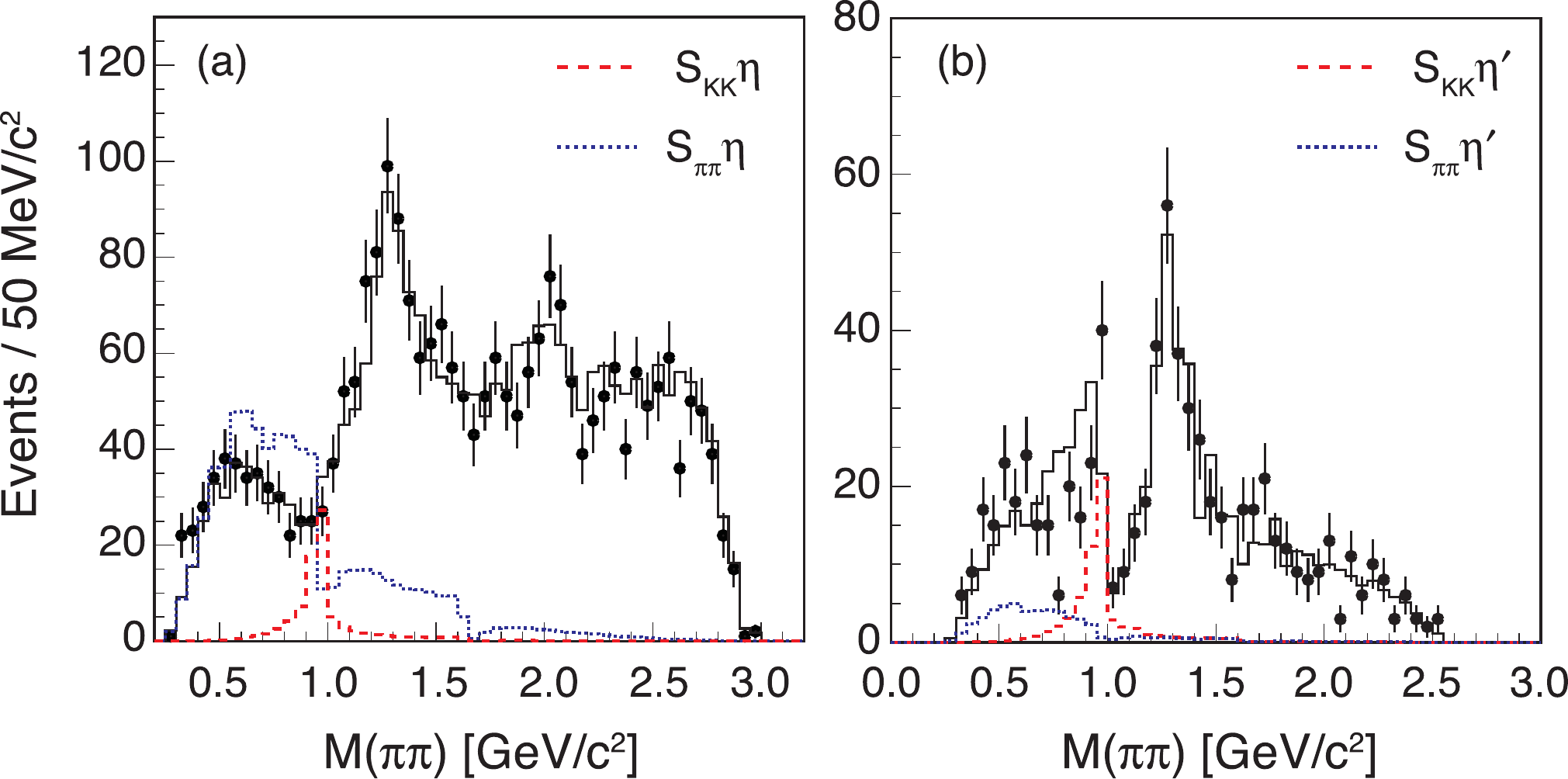}
\caption{Comparison of the $\pi\pi$ spectra and the $\pi\pi$ $S$-wave contributions to the 
$\chin\rar\eta\pip\pim$ (a) and
$\chin\rar\eta'\pip\pim$ (b) decays.  
\label{fig:m23_comp}
}
\end{figure*}

Since the $\eta$ and $\eta'$ have a well known composition in both the quark and SU(3) flavor bases, it may be valuable to examine ratios of branching fractions for $\chi_{c1}$ decay to $\pi\pi$ isobars and an $\eta$ or $\eta'$.  Figure~\ref{fig:m23_comp} shows a comparison of the $\pi\pi$ spectra for the fits when the recoil particle is an (a) $\eta$ and (b) $\eta'$.  The two components of $\pi\pi$ $S$-wave corresponding to $KK\to\pi\pi$ ($S_{KK}$) and $\pi\pi\to\pi\pi$ ($S_{\pi\pi}$) scattering have been highlighted.  There are qualitative differences in the two $\pi\pi$ $S$-waves.  Most notably, in the $\eta'$ recoil case, the $S_{KK}$ amplitude dominates the $S_{\pi\pi}$ amplitude producing a peak near the $f_0(980)$.  However, for the $\eta$ recoil case, the interference between $S_{KK}$ and $S_{\pi\pi}$ produces a dip near $f_0(980)$, which is more clearly visible in the coherent sum of these amplitudes depicted in Fig.~\ref{fig:mfits}(b).  A compilation of branching ratios for all $\pi\pi$ isobars is listed in Table~\ref{tab:pipi_rat}.

\begin{table}
\caption{\label{tab:pipi_rat}
Branching ratios of the $\chi_{c1}$ to various $\pi\pi$ states recoiling against $\eta$ and $\eta^\prime$.The errors are statistical and systematic, respectively.  Limits are set at the 90\% confidence level.  Correlations in errors on the individual fit fractions have been accounted for in the error on the ratios.
}
\medskip
\begin{center}
\begin{tabular}{lc}
\hline
\hline
  $\chi_{c1}$ Branching Ratio\hspace{1.5cm}  &  Value \\
\hline            
$(\pi\pi)_{S}\etaprime$/$(\pi\pi)_{S}\eta$ 	&	$	0.37	\pm	0.05	\pm	0.06	$	\\
\hspace{0.20cm} $S_{KK}\etaprime$/$S_{KK}\eta$ 	&	$	1.03	\pm	0.28	\pm	0.17	$	\\
\hspace{0.20cm} $S_{\pi\pi}\etaprime$/$S_{\pi\pi}\eta$ 	&	$	0.11	\pm	0.04	\pm	0.04	$	\\
$\ftw\etaprime$/$\ftw\eta$ 	&	$	1.39	\pm	0.20	\pm	0.11	$	\\
$\ffo\etaprime$/$\ffo\eta$ 	&	$	<		0.20			$	\\
\hline\hline
\end{tabular}
\end{center}
\end{table}

Table~\ref{tab:etapi_rat} lists the branching ratios for $\eta\pi$ and $\eta^\prime\pi$ decay channels of various isobars.  The fact that the $a_0(980)$ is below $\eta'\pi$ threshold undoubtedly contributes to the small  $\eta'\pi$ to $\eta\pi$ branching ratio.  The small $a_2(1320)\pi$ amplitude in our baseline fit yields an upper limit that is consistent with previous measurements:  ${\cal B} (\atw\to\etaprime\pi) / {\cal B} (\atw\to\eta\pi) = 0.037 \pm 0.006$~\cite{PDG10}.  Finally we can place a lower limit on $\mathcal{B}(\pi_1\to\eta'\pi)/\mathcal{B}(\pi_1\to\eta\pi)$ by adding a $\pi_1\pi$ amplitude to our fit to the $\eta\pi\pi$ data.  This limit indicates that the $\eta'\pi$ partial width of the $\pi_1$ is much larger than the $\eta\pi$ partial width.

\begin{table}
\caption{\label{tab:etapi_rat}
Branching ratios and limits for the $\aze$, $\atw$, and $\pione$ 
decays into the $\etaprime\pi$ and $\eta\pi$ decay modes.  The errors are statistical and systematic, respectively, and the limits are set at the 90\% confidence level.  Correlations in errors on the individual fit fractions have been accounted for in the error on the ratios.}
\medskip
\begin{center}
\begin{tabular}{lc}
\hline
\hline
  Branching Ratio   &  Value    \\
\hline            
 $\aze\to\etaprime\pi/\aze\to\eta\pi$	&	$	0.064	\pm	0.014	\pm	0.014	$	\\
$\atw\to\etaprime\pi / \atw\to\eta\pi$	&	$	<	0.065				$	\\
$\pione\to\etaprime\pi / \pione\to\eta\pi$	&	$	>	9.1				$	\\
\hline \hline
\end{tabular}
\end{center}
\end{table}


\section{\label{Systematics}Systematic errors}

In general, the systematic uncertainties on the results of this analysis can be classified into two broad categories:  those that affect the measurement of the fit fractions ($\mathcal{F}$) and those that affect the measurement of the total branching fraction for $\chin\rar\eorep\pip\pim$.  Table~\ref{tab:sys_err} presents a summary of the systematic errors on these quantities.  Correlations are considered when assigning systematic errors to branching ratios and products.

\begin{table*}[ht]
 \caption{\label{tab:sys_err}
 Fractional systematic errors in percent on the measurements of $\chin\rar\eorep\pip\pim$ branching fractions
and amplitude contributions. See text for explanations.
}
\medskip
\begin{center}
\begin{tabular}{lcccccc|c}																					
\hline																					
\hline																					
Decay Mode	&  ~~$N_{\psi(2S)}$~~	&	~~$\epsilon_{\pi\pm}$~~	&	~~$\epsilon_{\gamma}$~~	&	 ~~$E1/M2$~~	&	~~Bkg.~~ 	&	 ~~$T_{\alpha}(s)$~~ &		Total Systematic	\\			
\hline
$\eta\pip\pim$ 	&	2.0	&	0.8	&	3.4	&	0.7	&	1.2	&	-	&		4.3	\\	
\hspace{0.20cm} $\aze\pi$	&	-	&	-	&	-	&	0.2	&	0.4	&	1.3	&	4.4		\\	
\hspace{0.20cm} $\atw\pi$ 	&	-	&	-	&	-	&	0.9	&	7.1	&	7.6	&	11.2		\\	
\hspace{0.20cm} $(\pip\pim)_{S}\eta$	&	-	&	-	&	-	&	0.5	&	3.4	&	10.7	&	11.9		\\	
\hspace{0.50cm} $\sppz\eta$ 	&	-	&	-	&	-	&	2.1	&	2.8	&	45.7	&	46.1		\\	
\hspace{0.50cm} $\sppo\eta$	&	-	&	-	&	-	&	2.6	&	7.3	&	42.2	&	43.2		\\	
\hspace{0.50cm} $\skk\eta$	&	-	&	-	&	-	&	1.3	&	4.2	&	11.6	&	13.1		\\	
\hspace{0.20cm} $\ftw\eta$ 	&	-	&	-	&	-	&	0.9	&	5.9	&	6.2	&	9.6		\\	
\hspace{0.20cm} $\ffo\eta$	&	-	&	-	&	-	&	1.0	&	27.8	&	13.4	&	31.2		\\	
\hline																				
$\etaprime\pip\pim$	&	2.0	&	1.2	&	2.9	&	0.6	&	2.0	&	-	&		4.3	\\	
\hspace{0.20cm} $\aze\pi$	&	-	&	-	&	-	&	1.0	&	9.4	&	13.1	&	16.7		\\	
\hspace{0.20cm} $(\pip\pim)_{S}\eta$	&	-	&	-	&	-	&	1.6	&	4.6	&	2.6	&	7.0		\\	
\hspace{0.50cm} $\sppz\etaprime$ 	&	-	&	-	&	-	&	1.3	&	21.3	&	24.1	&	32.5		\\	
\hspace{0.50cm} $\skk\etaprime$	&	-	&	-	&	-	&	0.8	&	8.5	&	12.4	&	15.6		\\	
\hspace{0.20cm} $\ftw\etaprime$ 	&	-	&	-	&	-	&	0.5	&	3.2	&	5.4	&	7.5		\\	
\hspace{0.20cm} $\pione\pi$ 	&	-	&	-	&	-	&	2.7	&	10.3	&	18	&	21.4		\\	
\hline\hline

\end{tabular}
\end{center}
\end{table*}

The total branching fraction of $\chin\rar\eorep\pip\pim$ is affected by uncertainties in the Monte Carlo model of the track and photon efficiency, $\epsilon_{\pi\pim}$ and $\epsilon_\gamma$.  We assume a 0.3\% systematic error for each track and 1\% systematic error each photon.  The total error is obtained by assigning a systematic error to each decay mode of the $\eta$ or $\eta^\prime$ and then constructing a weighted average of these individual errors where the weights are given by the product of the branching fraction and detection efficiency, {\it i.e.}, a weight proportional to the number of observed events, for each mode.  Assuming that the systematic error is not dependent on location of the event in the $\eorep\pip\pim$ phase space, these errors cancel in the determination of the fit fractions.

Background events have the potential to affect our analysis in two different ways.  First, our computation of the purity $p$ of each sample obtained from fits to the spectra shown in Fig.~\ref{fig:mChic1} may be subject to systematic bias that would affect the measurement of $\mathcal{B}(\chin\rar\eorep\pip\pim)$.  Second, because the background level is small and its angular distributions are difficult to characterize, our fit does not include a background amplitude.  It is assumed that the background distributes itself among the various amplitudes thereby leaving the fit fractions unchanged.  We test this assumption and also probe the stability of our total branching fraction measurement by relaxing event selection cuts and repeating the analysis.  Specifically, we relax the $\chidof$ requirement; widen the invariant mass regions used to select $\eta$, $\eta^\prime$, and $\chi_{c1}$; introduce some $J/\psi$ background by reducing the effectiveness of the $J/\psi$ suppression criteria; and enhance the probability that photons reconstructed in the event are actually decay products of $\pi^0$ by reducing the effectiveness of the $\pi^0$ veto requirement.  We take the largest deviation from our baseline analysis as the systematic error due to background.

The construction of our amplitudes assumes the radiative transition $\psi(2S)\to\gamma\chi_{c1}$ is purely electric dipole $E1$.  However, the contribution of magnetic quadrupole $M2$ amplitude has been 
measured by CLEO~\cite{CLEO506} to be $(2.76\pm0.76)\%$.   This slightly affects the angular distribution of the radiated photon and the polarization of the $\chi_{c1}$.  We quantify the uncertainty due to this assumption by repeating the analysis using the measured CLEO value and assigning the deviation from our baseline analysis as the systematic error.  The dominant effect on the measurement of $\mathcal{B}(\chin\rar\eorep\pip\pim)$ is due to the change in detection efficiency of the radiated photon, while altering the polarization of the $\chi_{c1}$ affects the fitted values of $\mathcal{F}$.  While these two effects are independent, their source is fully correlated, and we take this into account when obtaining product branching fractions.

The choice of parametrization for the two-body dynamics $T_\alpha(s)$ has the potential to systematically bias the results.  While the error in the total branching fraction due to such variations is negligible, the individual fit fractions can be strongly affected by variations in $T_\alpha(s)$.  We vary, individually, the mass and width of the $\atw$ and $\ftw$ within one standard deviation as tabulated by the PDG~\cite{PDG10} and repeat the analysis.  We also vary the parameters of the $a_0(980)$ according to the uncertainties listed in Table~\ref{tab:xlow_a0par}.  The mass and width parameters of the $\pi_1(1600)$ and $\ffo$, which are fixed in our baseline analysis to their best-fit values, are also varied by one standard deviation, and the analysis is repeated. Finally, we test the sensitivity of our results to the parametrization of the $\pi\pi~S$-wave by scaling the magnitude of this distribution by a value that ranges from unity to 1.2 or 0.8 linearly with $s$.   That is, we try various random linear changes in the shape at the 20\% level.  In addition, we vary the parameter $s_0$ in Eq.~(\ref{eq:Zsubs}).  The largest deviation from our baseline fits within this set of variations is taken as the systematic uncertainty due to the amplitude parametrization.

Finally, we must use as inputs the number of $\psi(2S)$ events in the data sample and branching fractions for the relevant $\eta$, $\eta^\prime$, and $\psi(2S)$ decays.  The number of $\psi(2S)$ decays is known to a precision of 2\%.  The other branching fractions and their errors are taken from the PDG review~\cite{PDG10} and listed as a separate, external systematic error for each measured quantitiy.


\section{\label{Conclude}Conclusions}

We present an analysis of $\psi(2S)\rar\gamma\chin\rar\gamma\eorep\pip\pim$ decays in which we study the production of various $\eta^{(\prime)}\pi$ and $\pi\pi$ intermediate states.  Both channels exhibit a signal purity of at least 95\% and the majority of $\eta$ and $\eta^\prime$ decay modes are utilized in the analysis.  The contributions from various quasi two-body decays are extracted utilizing an unbinned maximum likelihood fit that spans the phase space of relevant kinematic variables needed to describe the decay.  Two-body interactions in our model are parametrized by both Breit-Wigner and Flatt\'e distributions, and we utilize a parametrization of the $\pi\pi~S$-wave interactions that is based on scattering data.  The results presented here supersede those previously presented by the CLEO Collaboration~\cite{CLEOCHI}.

We find evidence for an exotic $\eta^\prime\pi$ $P$-wave scattering amplitude at the level of 4 standard deviations under a wide variety of model variations.  If we parametrize this amplitude as a Breit-Wigner resonance we obtain a mass and width that is consistent with the $\pi_1(1600)$ state reported in the literature.  While the best description of the data is achieved with a resonant $\pi_1\pi$ amplitude, it is impossible to exclude other mechanisms that contribute to the $\eta^\prime\pi$ $P$-wave amplitude.  In addition, the $\chi_{cJ}\to\eta'\pip\pim$ data provide the first direct evidence for the decay of $a_0(980)\to\eta'\pi$.  We measure the ratio of branching fractions for the $\eta'\pi$ and $\eta\pi$ decay channels of the $a_0(980)$.  The $\chi_{c1}\to\eta\pip\pim$ data allow us to extract the parameters of the $a_0(980)$ lineshape in the context of a three-channel Flatt\'e distribution.  Finally we compare the $\pi\pi$ system when it is produced against an $\eta$ to that produced against an $\eta'$.  Our model for the $\pi\pi$ $S$-wave interactions suggests that production via $\pi\pi\to\pi\pi$ $S$-wave scattering is suppressed with respect to $KK\to\pi\pi$ scattering when the system recoils against the $\eta'$.  We also extract similar branching ratios for the other $\pi\pi$ resonances used in the fit, the $\ftw$ and $\ffo$.

We gratefully acknowledge the effort of the CESR staff
in providing us with excellent luminosity and running conditions.
D.~Cronin-Hennessy thanks the A.P.~Sloan Foundation.
We would like to thank P.~Guo and H.~Matevosyan for their helpful discussions.
This work was supported by
the National Science Foundation,
the U.S. Department of Energy,
the Natural Sciences and Engineering Research Council of Canada, and
the U.K. Science and Technology Facilities Council.  Additional support for computational hardware was provided by the Indiana University Office of the Vice Provost for Research.

\end{document}